\documentclass[pra,aps,10pt,twocolumn,superscriptaddress]{revtex4}
\usepackage{amsmath}
\usepackage{amssymb}
\usepackage{amsthm}
\usepackage{amsfonts}
\usepackage{enumerate}
\usepackage{latexsym}
\usepackage{bm}
\usepackage{graphicx}
\usepackage{subfigure}
\usepackage{booktabs}
\usepackage[dvipsnames]{xcolor}
\usepackage[colorlinks]{hyperref}
\usepackage{multirow}
\usepackage[normalem]{ulem}

\definecolor{darkgreen}{RGB}{0,128,0}

\definecolor{orange}{RGB}{255,139,61}

\newcommand{\beq}{\begin{equation}}
\newcommand{\eneq}{\end{equation}}

\newcommand{\ket}[1]{| #1 \rangle}

\input{epsf}

\begin{document}

\title{Probing weak dipole-dipole interaction using phase-modulated non-linear spectroscopy}
\author{Zeng-Zhao Li} 
\affiliation{Max Planck Institute for the Physics of Complex Systems,  N\"othnitzer Strasse 38, 01187 Dresden, Germany}
\author{Lukas Bruder}
\affiliation{Physics Department, University of Freiburg, Hermann-Herder-Str. 3, D-79104 Freiburg, Germany}
\author{Frank Stienkemeier}
\affiliation{Physics Department, University of Freiburg, Hermann-Herder-Str. 3, D-79104 Freiburg, Germany}
\author{Alexander Eisfeld}
\affiliation{Max Planck Institute for the Physics of Complex Systems,  N\"othnitzer Strasse 38, 01187 Dresden, Germany}
\email{eisfeld@pks.mpg.de}

\begin{abstract}

 Phase-modulated non-linear spectroscopy with higher harmonic demodulation has recently been suggested to provide information on many-body excitations.
 In the present work we theoretically investigate the application of this method to infer the interaction strength between two particles that interact via weak dipole-dipole interaction.
To this end we use full numerical solution of the Schr\"odinger equation with time-dependent pulses.  For interpretation purpose we also derive analytical expressions in perturbation theory.
 We find one can detect dipole-dipole interaction via peak intensities (in contrast to line-shifts which typically are used in conventional spectroscopy).
We provide a detailed study on the dependence of these intensities on the parameters of the laser pulse and the dipole-dipole interaction strength.
Interestingly, we find that there is a phase between the  first and second harmonic demodulated signal, whose value depends on the sign of the dipole-dipole interaction.

\end{abstract}

\date{\today}
%\pacs{32.80.Qk,32.80.Wr}
% 32.80.Wr	Other multiphoton processes
% 32.80.Qk	Coherent control of atomic interactions with photons
\maketitle

\section{Introduction \label{sec:intro}}

The study of many-body effects in weakly interacting ensembles of particles poses challenges to experiments.
In many cases strong single-particle signals or broadening effects mask the weak collective signals. 
Recently, coherent time-resolved non-linear spectroscopy introduced a very sensitive approach to probe weak many-body effects with double-quantum two-dimensional (2D) spectroscopy\,\cite{StoneCundiffNelson09Science,KaraiskajCundiff10PRL,TurnerNelson10Nature,DaiCundiff12PRL,GaoCundiffLi16} where the two-body response is background-free detected, hence the mere presence of a double-quantum signal indicates the existence of inter-particle interactions in the system. 
Double quantum coherence measurements in 2D spectroscopy are very useful to detect weak interactions. However, most 2D schemes do not provide sufficient sensitivity to investigate dilute samples.

A multidimensional spectroscopy approach suitable for highly dilute samples is based on phase modulation (PM)~\cite{TekavecMarcus07JCP}. 
This approach has been demonstrated in 2D electronic spectroscopy~\cite{TekavecMarcus07JCP} but also in electronic wave packet interferometry~\cite{TekavecMarcus06JCP}. In the latter scheme, one records the signal from an observable (such as fluorescence intensity, photocurrent or photoionization~\cite{TekavecMarcus06JCP,NardinCundiff13,BruderStienkemeier2015PCCP}) emitted by the particles subject to two short laser pulses with a specific time-delay and a slowly modulated relative phase. 
Demodulation with respect to this relative phase and a subsequent Fourier transformation with respect to the time-delay yield a complex-valued spectrum. 
 By using a variant of this technique, in Ref.~\onlinecite{BruderStienkemeier15PRA} the demodulation has been performed by using higher harmonics of the reference signal.

In the present work we aim at  understanding the effect of long-range dipole-dipole interaction on the spectra resulting from higher order demodulation. 
To this end we investigate a minimal model consisting of two  particles which can interact via transition dipole-dipole interaction. 
The model that we use is in particular related to interacting dye molecules which have a parallel arrangement.
To infer the dipole-dipole interaction in such systems is of considerable interest (see e.g.~Refs.~\cite{WeSt03_125201_,FiSeEn08_12858_,Ei07_321_,GuZuCh09_154302_,WoMo09_15747_,DuNaPr15_072002_}).
The model of our measurement results effectively in a one-body measurement operator (number of detected photons).
Our work will thus provide reference for the predictions from such a single-particle operator.

In our theoretical investigations we will on the one hand use a non-perturbative fully numerical approach, on the other hand we will derive analytical formulas valid if a perturbation treatment with respect to the light-matter interaction is applicable.
In our numerical calculations we aim at following the experimental procedure as close as possible.

The structure of this paper is as follows. In Sec.~\ref{sec:model}, we introduce a model of multiple particles that interact with laser fields and also a  theoretical background of the phase demodulation technique used to reveal collective behaviors of particles.  
Taking two particles as a typical example, in Sec.~\ref{sec:simulation} we present the fully numerical simulation results for Gaussian pulses. 
In Sec.~\ref{sec:analytics}, basing on perturbation theory, we develop analytical formulas of phase-(de)modulated fluorescence signal responsible for that measured in the experiment and present analytical results of signals for rectangular pulses in the appendix. 
Discussions and conclusions are finally given in Sec.~\ref{sec:conclusion}. 
In a supporting information we provide details about our numerical and analytical procedures.
Throughout the paper we set $\hbar=1$.

\section{The model \label{sec:model}}
\begin{figure}[t]
\centering
  \includegraphics[width=.95\columnwidth]{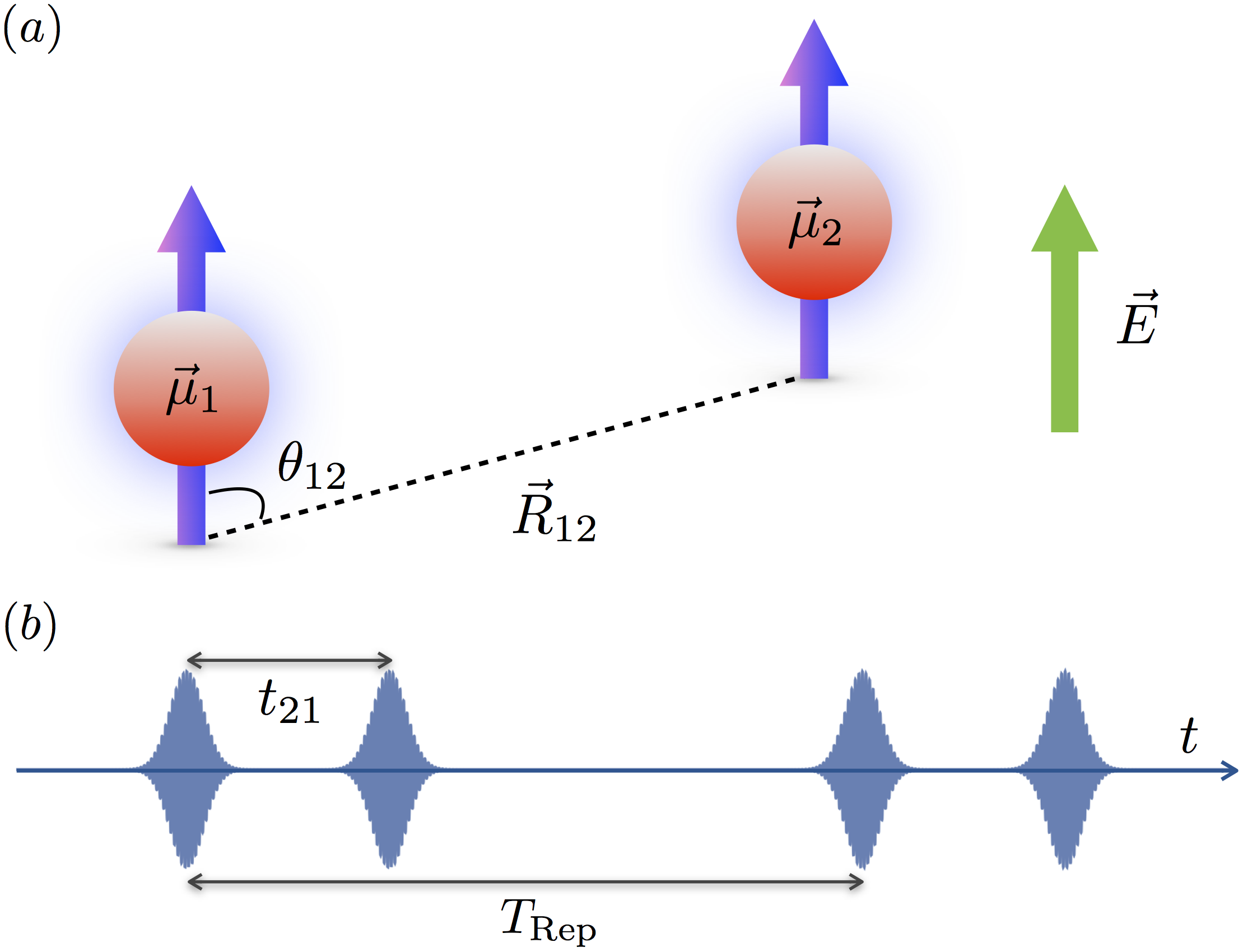} 
\caption{(color online)  (a) Arrangement of the particles and the laser polarization. (b)  Illustration of two pulse pairs where $T_{\mathrm{Rep}}$ is the repetition time between pairs and $t_{21}$ is the pulse delay between the pulses within each pair.}
\label{fig:schematic_system+pulse}
\end{figure}

\subsection{System and interaction with laser pulses}
We consider an ensemble of particles that do not move and that are so far apart that only long-range interaction of the dipole-dipole type has to be taken into account.
The Hamiltonian for $N$ particles coupled to a pulsed laser field reads
\begin{eqnarray}
H&=&\sum_{n=1}^N H_{n} +\sum_{nn'}V_{nn'} +\sum_{n=1}^N H_{n}^{\rm int}(t). \label{eq:total_H}
\end{eqnarray}
The Hamiltonian $H_n$ describes the individual particles, $V_{nn'}$ is the dipole-dipole interaction between particle $n$ and $n'$, and $H_n^{\rm int}$ is the interaction of particle $n$ with the electromagnetic field.
 In the following subsections we discuss the various parts of the Hamiltonian above.
Throughout the work we focus on the case of two particles (i.e. $N=2$).
Our basic setup is sketched in Fig.~\ref{fig:schematic_system+pulse}.

\subsubsection{Individual particles $H_n$}

For each particle we consider three  energy eigenstates $\{|{\rm g}\rangle_n,|{\rm e}\rangle_n,|{\rm f}\rangle_n\}$ and write the Hamiltonian of particle $n$ as 
\begin{eqnarray}
H_n&=&\epsilon_{\rm g}|{\rm g}\rangle_n\langle {\rm g}| +\epsilon_{\rm e}|{\rm e}\rangle_n\langle {\rm e}| +\epsilon_{\rm f}|{\rm f}\rangle_n\langle {\rm f}|.
\label{eq:free_atom}
\end{eqnarray}
Here $|{\rm g}\rangle$, $|{\rm e}\rangle$ and $|{\rm f}\rangle$ represent the ground, first and second excited states, respectively.
We assume for simplicity and to elucidate the basic effects most clearly that there are dipole allowed transitions only between the ground state and each of the excited states, i.e.\ between $\ket{{\rm g}}$ and $\ket{{\rm e}}$, and between $\ket{{\rm g}}$ and $\ket{{\rm f}}$.
The corresponding dipoles we denote by $\vec{\mu}_{{\rm ge}}^{(n)}$ and $\vec{\mu}_{{\rm gf}}^{(n)}$, respectively, and take them   to be parallel (and identical for the two particles).

\subsubsection{Dipole-dipole interaction $V_{nn'}$}

The basis states of two particles $n$ and $n'$ can be constructed as product states of single particles, namely $\{|{\rm gg}\rangle,|{\rm ge}\rangle,|{\rm eg}\rangle,|{\rm gf}\rangle,|{\rm fg}\rangle,|{\rm ee}\rangle,|{\rm ef}\rangle,|{\rm fe}\rangle,|{\rm ff}\rangle\}$, where $|\alpha\beta\rangle=|\alpha\rangle_i |\beta\rangle_j$ ($\alpha, \beta={\rm g}, {\rm e}, {\rm f}$). The dipole-dipole interaction between two particles is then written in terms of these basis states as \cite{May-Kuehn}
\begin{eqnarray}
V_{nn'}&=&V_{nn'}^{{\rm ee}}
|{\rm ge}\rangle\langle {\rm eg}|
+V_{nn'}^{{\rm ff}}
|{\rm gf}\rangle\langle {\rm fg}|
\notag \\
&&
+V_{nn'}^{{\rm ef}}(|{\rm ge}\rangle\langle {\rm fg}|
+|{\rm gf}\rangle\langle {\rm eg}|
) +H.c..
\label{eq:dipole_dipole}
\end{eqnarray}
For parallel transition dipoles (see Fig.~\ref{fig:schematic_system+pulse}) the interaction strengths are given by
\begin{eqnarray}
V_{nn'}^{\alpha\beta}&=&\frac{\mu_{g\alpha}^{(n)} \mu_{g\beta}^{(n')}}{R_{nn'}^3}(1-3\cos^2\theta_{nn'}),
\label{eq:dipole-dipole_ampl}
\end{eqnarray}
with $\vec{R}_{nn'}$ being the vector between the two point-dipoles ($R_{nn'}=|\vec{R}_{nn'}|$), $\theta_{nn'}$ is the angel between $\vec{R}_{nn'}$ and the dipole orientation. 
Note that depending on the angle $\theta_{nn'}$  the sign of the interaction can change from positive to negative.
To facilitate the analytical considerations we neglect off-resonant contributions coupling the two excited states via dipole-dipole interaction $V_{nn'}^{\rm ef}$. 
This approximation is good when the dipole-dipole interaction is much smaller than the energy separation between the two states (which is the case in which we are primarily interested in).

Note that from the definition Eq.~(\ref{eq:dipole-dipole_ampl}) the interaction $V$ depends on the distance between the two particles $R$ and on the angle $\theta$.
That means that for a given value of $V$ there exist many possible configurations $R$, $\theta$.
Note also, that for certain angles $\theta$ the interaction $V$ can be negative.

\begin{figure}[t]
\centering
  \includegraphics[width=.95\columnwidth]{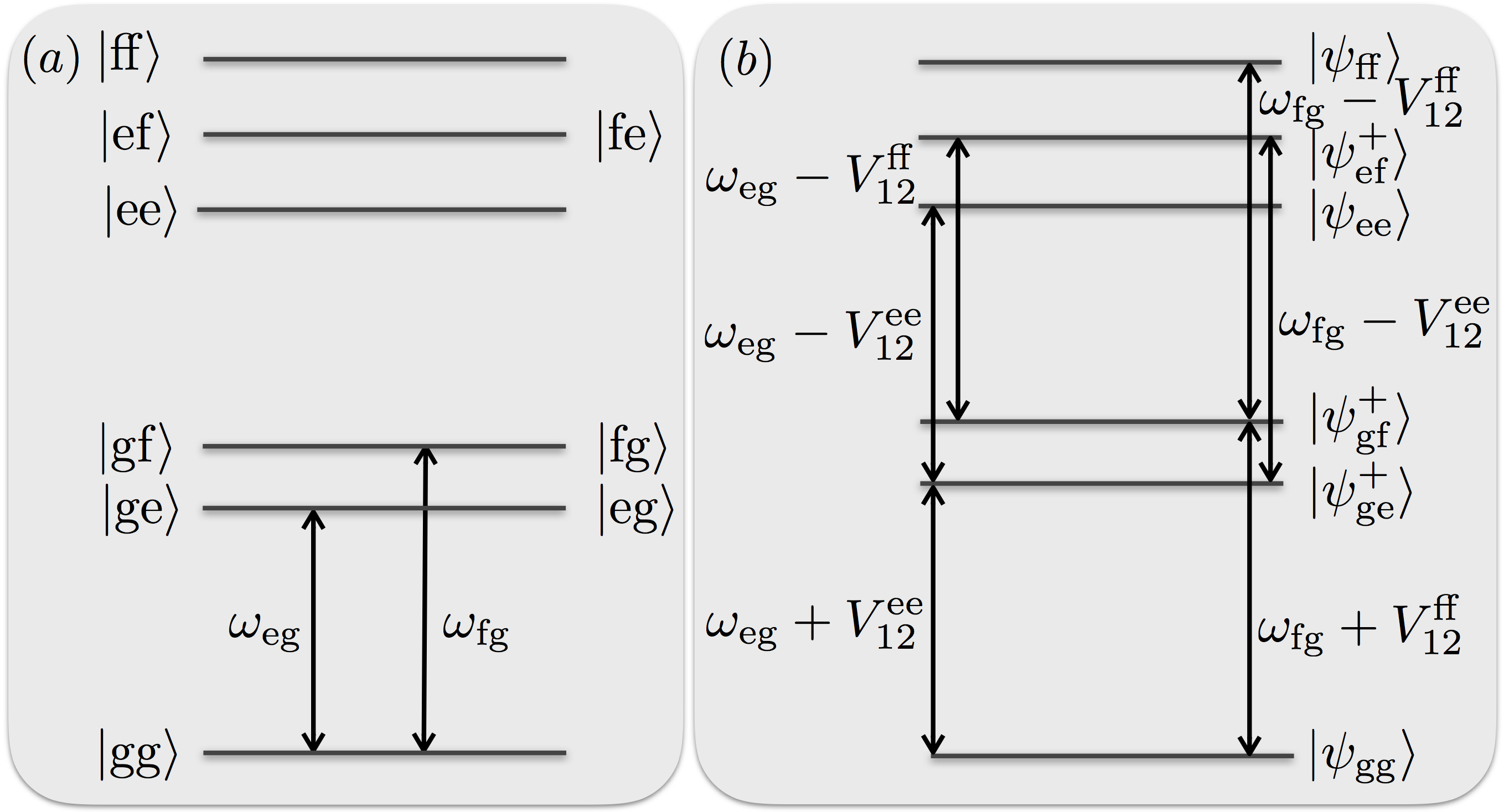} 
\begin{tabular}{|l|l|}
\hline
eigenstate & energy \\
\hline
$|\psi_{{\rm gg}}\rangle=|{\rm gg}\rangle$ & $ E_{{\rm gg}}=E_{\rm g}+E_{\rm g}  $\\
$|\psi_{{\rm ge}}^{\pm}\rangle=\frac{1}{\sqrt{2}}(|{\rm ge}\rangle\pm|{\rm eg}\rangle)$&$ E_{{\rm ge}}^{\pm}=E_{\rm g}+E_{\rm e}\pm V_{{\rm ee}}$ \\ 
$|\psi_{{\rm gf}}^{\pm}\rangle=\frac{1}{\sqrt{2}}(|{\rm gf}\rangle\pm|{\rm fg}\rangle)$&   $E_{{\rm gf}}^{\pm}=E_{\rm g}+E_{\rm e}\pm V_{{\rm ff}}$  \\ 
$|\psi_{{\rm ee}}\rangle=|{\rm ee}\rangle $ & $E_{\rm ee}= 2 E_{\rm e}$  \\
$|\psi_{{\rm ef}}^{\pm}\rangle=\frac{1}{\sqrt{2}}(|{\rm ef}\rangle\pm|{\rm fe}\rangle)$ &    $E_{{\rm gf}}^{\pm}=E_{\rm e}+E_{\rm f}$    \\ 
$|\psi_{{\rm ff}}\rangle=|{\rm ff}\rangle$ & $E_{\rm ff}= 2 E_{\rm f}$  \\
\hline
\end{tabular}
\caption{(color online) Schematic diagram of collective energy levels of two particles (a) without and (b) with  dipole-dipole interaction (only symmetric states are shown).
The arrows indicate dipole-allowed transitions. The corresponding energy-differences are also indicated.
Note that we ignore direct interactions between two excited particles (like van der Waals interactions) which would lead e.g.\ to additional level shifts. 
In the table the collective eigenstates and the corresponding energies are specified.
 }
\label{fig:schematic_diagram}
\end{figure}

\noindent
{\it Two particles: Eigenenergies and eigenstates:}
For two particles the Hamiltonian Eq.~(\ref{eq:total_H}) becomes  $H_1+H_2+V_{12}$ where $H_n$ and $V_{12}$ are given by Eqs.~(\ref{eq:free_atom}) and (\ref{eq:dipole_dipole}) respectively, where  $V_{12}^{\rm ee}$ and $V_{12}^{\rm ff}$ will be abbreviated by $V_{\rm ee}$ and $V_{\rm ff}$ in the following.
For the interpretation of the results it is convenient to use the eigenstates and eigenenergies. 
The energy levels without and with the dipole-dipole interaction are shown in Figs.~\ref{fig:schematic_diagram}(a) and (b) respectively.

The symmetric states, i.e. $|\psi_{{\rm gg}}\rangle$, $|\psi_{{\rm ge}}^+\rangle$, $|\psi_{{\rm gf}}^+\rangle$, $|\psi_{{\rm ee}}\rangle$, $|\psi_{{\rm ef}}^+\rangle$, and $|\psi_{{\rm ff}}\rangle$ 
span one subspace of the Hilbert space, while the other subspace is spanned by antisymmetric states, i.e. $|\psi_{{\rm ge}}^{-}\rangle$, $|\psi_{{\rm gf}}^{-}\rangle$, and $|\psi_{{\rm ef}}^{-}\rangle$. 
For the definition of these states see Fig.~\ref{fig:schematic_diagram}.

\subsubsection{Particle-field interaction $H_n^{{\rm int}}$}

The interaction between each particle and an electromagnetic field  is described  by
\begin{eqnarray}
H_{n}^{\rm int}(t) &=& \Omega_n^{{\rm e}}(t)
|{\rm g}\rangle_n\langle {\rm e}|
+\Omega_n^{{\rm f}}(t) 
|{\rm g}\rangle_n\langle {\rm f}|
+H.c.,
\label{eq:atom_field}
\end{eqnarray}
with time-dependent coupling strengths $\Omega_n^{{\rm e}}(t)=-\vec{\mu}_{{\rm ge}}^{(n)} \vec{E}(t)$ and  $\Omega_n^{{\rm f}}(t)=-\vec{\mu}_{{\rm gf}}^{(n)} \vec{E}(t)$ with $\vec{E}(t)$ the laser pulse field.
Again for simplicity we take   $\vec{E}(t)$ to be identical for the two particles and parallel to the transition dipoles of the particles (see Fig.~\ref{fig:schematic_system+pulse}(a)).
For identical particles in our consideration, i.e. $\mu_{{\rm ge}}^{(n)}=\mu_{\rm e}$ and $\mu_{{\rm gf}}^{(n)}=\mu_{\rm f}$, the coupling strengths then become $\Omega_n^{{\rm e}}(t)=\Omega^{{\rm e}}(t)$ and $\Omega_n^{{\rm f}}(t)=\Omega^{{\rm f}}(t)$.

\subsection{Modelling the measurement procedure}
In the considered PM pump-probe experiment, two phase-modulated femtosecond (fs) laser pulses excite the system. 
Fluorescence intensity is recorded via time-integrated detection in a small angle perpendicular to the excitation laser propagation direction.
Thereby it is not distinguished between fluorescence emitted from a many-particle or single-particle population. 
The many-body information is encoded in the time evolution between the two pulses resulting in a time-dependence of the fluorescence yield.

Each pulse is tagged with an individual phase signature stemming from the phase modulation, allowing for filtering of certain contributions of the final signal, as described below. 
This phase signature is manifested in a linear phase sweep $\phi_j(t)=\Omega_j t + \phi_j^{(0)}$ imparted by acousto-optic devices. 
The laser field consisting of a train of pulse pairs and illustrated in Fig.~\ref{fig:schematic_system+pulse}(b) is hence expressed as
\begin{eqnarray}
E(t) &=& \sum_{m=0}^M \sum_{j=1}^2 A_j(t-t_j-\tau_m)\cos[\omega_j(t-t_j-\tau_m) \notag \\
&& +\Omega_j t+\phi_j^{(0)}],
\label{eq:pulsefield}
\end{eqnarray}
where $A_j(t-t_j-\tau_m)$ is the pulse envelope function.
Here $t_j$ is the arrival time of the $j$th pulse in a pulse pair  and $\tau_m=mT_{\mathrm{Rep}}$ with $m$ being the number of pulse pairs and $T_{\mathrm{Rep}}$ the pulse repetition time. 
In Eq.~(\ref{eq:pulsefield}), $\omega_j$ is the laser central frequency, $\phi_j^{(0)}$ is the phase offset, and the term $\Omega_j t$ describes the linear phase sweep due to the pulses passing individual acousto-optic modulators. 

The PM $\phi_j(t)$ results in a slow intensity modulation of the fluorescence which allows one to use lock-in techniques after recording the fluorescence intensity with a photodetector. 
Lock-in detection can even be used to filter the fluorescence signal with respect to harmonics of the modulation frequency~\cite{TianWarren2002OptLett,KaKrMa16_015504_,BruderStienkemeier15PRA}. 
In this way individual many-body responses are separated into different signal channels.  
Fourier-transforming the final demodulated signals yields complex-valued frequency spectra.

We assume that the signal produced by the photodetector is proportional to the number of photons emitted by the system which we take to be  proportional to the number of excited particles after each pulse pair, ignoring emission during the short time-scale of the pulses. 
We take $T_{\mathrm{Rep}}$  much larger than the radiation life time of the particles.
Therefore, we can consider each pulse pair individually. 
We use the state after this pulse pair (at time $T_{\rm F}$ sufficiently large) to calculate the number of photons that will be emitted %during 
after the pulse pair: 
\begin{eqnarray}
S_{\rm Fluor}(t_{21},\tau_m) &=& \mathrm{Tr}\{P_{\rm Fluor}\rho(T_{\rm F}; t_{21},\tau_m)\} \notag \\
&=&\langle \Psi(T_{\rm F})| P_{\rm Fluor} |\Psi(T_{\rm F})\rangle ,
\label{eq:mod_signal_defin}
\end{eqnarray}
 where $\rho(t;t_{21},\tau_m )$ is the density matrix at time t, obtained for the pulse delay $t_{21}=t_2-t_1$ and the modulation time $\tau_m$
and
\begin{equation}
 P_{\rm Fluor}=\sum_n (|{\rm e}\rangle_n\langle {\rm e}|+|{\rm f}\rangle_n\langle {\rm f}|). 
\end{equation}

\subsubsection{Signal demodulation}

The signal Eq.~(\ref{eq:mod_signal_defin}) is modulated as a function of $\tau_m$ by the frequency difference $\Omega_{21}=\Omega_2 - \Omega_1$. 
In the experiment harmonic demodulation with a lock-in amplifier as described in Ref.~\onlinecite{BruderStienkemeier15PRA} is applied. 
Basically, the lock-in amplifier multiplies the signal with a known reference waveform and applies subsequently a low-pass filter, thus the demodulated output signal can be described by: 
\begin{eqnarray}
\tilde{S}_{\rm Fluor}(t_{21},\kappa) &=& \frac{1}{\tau_{\rm LI}} \int_0^{\infty} d\tau_m S_{\rm F1uor}(t_{21},\tau_m) \notag \\
&& \times S_{\mathrm{ref}} (t_{21},\tau_m,\kappa) e^{-\frac{\tau_m}{\tau_{\rm LI}}}, \label{eq:demodulation}
\end{eqnarray}
where $\tau_{\rm LI}$ is the lock-in time, $\kappa$ is the demodulation order, and $S_{\mathrm{ref}} (t_{21},\tau_m,\kappa)$ is a complex-valued reference signal given by
\begin{eqnarray}
S_{\mathrm{ref}} (t_{21},\tau_m,\kappa) \!\!\!&=&\!\!\! e^{i(\kappa\omega_{\rm M} t_{21}-\kappa\Omega_{21} \tau_m-\kappa\phi_{21}^{(0)})},
\end{eqnarray}
where $\phi_{21}^{(0)}=\phi_2^{(0)}-\phi_1^{(0)}$. 
In this work we focus on the first ($\kappa=1$) and second ($\kappa=2$) harmonic demodulation. 
We will abbreviate them as 1HD and 2HD, respectively.
Note that for sufficiently large $\tau_{\rm LI}$ the demodulation Eq.~(\ref{eq:demodulation}) essentially results in extracting the frequency components with $\kappa\Omega_{21} \tau_m$ out of the signal and shifting the frequencies of the resulting function by $\kappa \omega_{\rm M}$.
In our theoretical considerations we will use $\omega_{\rm M}=0$.

For visualization it is convenient to perform a Fourier-transformation with respect to $t_{21}$.
The resulting (complex-valued) quantity
 \begin{equation}
S(\omega,\kappa)= \frac{1}{\sqrt{2\pi}} \int_{0}^{\infty} dt_{21} \, e^{-i \omega t_{21}} \tilde{S}_{\rm Fluor}(t_{21},\kappa)
\label{eq:Swk}
\end{equation}
will be denoted as the spectrum in the $\kappa$th order of demodulation.

\section{Fully numerical simulation \label{sec:simulation}}

\begin{table}[b]
\caption{Parameters that are fixed in all numerical calculations shown.
All  energis are expressed in units of $\omega_{0}$. Time is in units of $1/\omega_{0}$.}
\centering
\begin{tabular}{|c|c|c|c|c|c|c|c|c|}
\hline\hline
Parameters  & $\omega_{\rm fe}$ &$\mu_{\rm e} $ & $\mu_{\rm f}$ & $\omega_{\rm M}$ & $\phi_{21}^{(0)}$ & $m$ &$\Omega_{21}$ & $T_{\rm Rep}$\\ [0.5ex]
\hline
Values  & $0.05$ & $0.75$ & $1.054$ & $0$ & $0$ & $1000$ &$2\pi \times 10^{-3} $& $1$ \\ [1ex]
\hline\hline
\end{tabular}
\label{table:pars}
\end{table}

\begin{figure}[t]
 \includegraphics[width=.98\columnwidth]{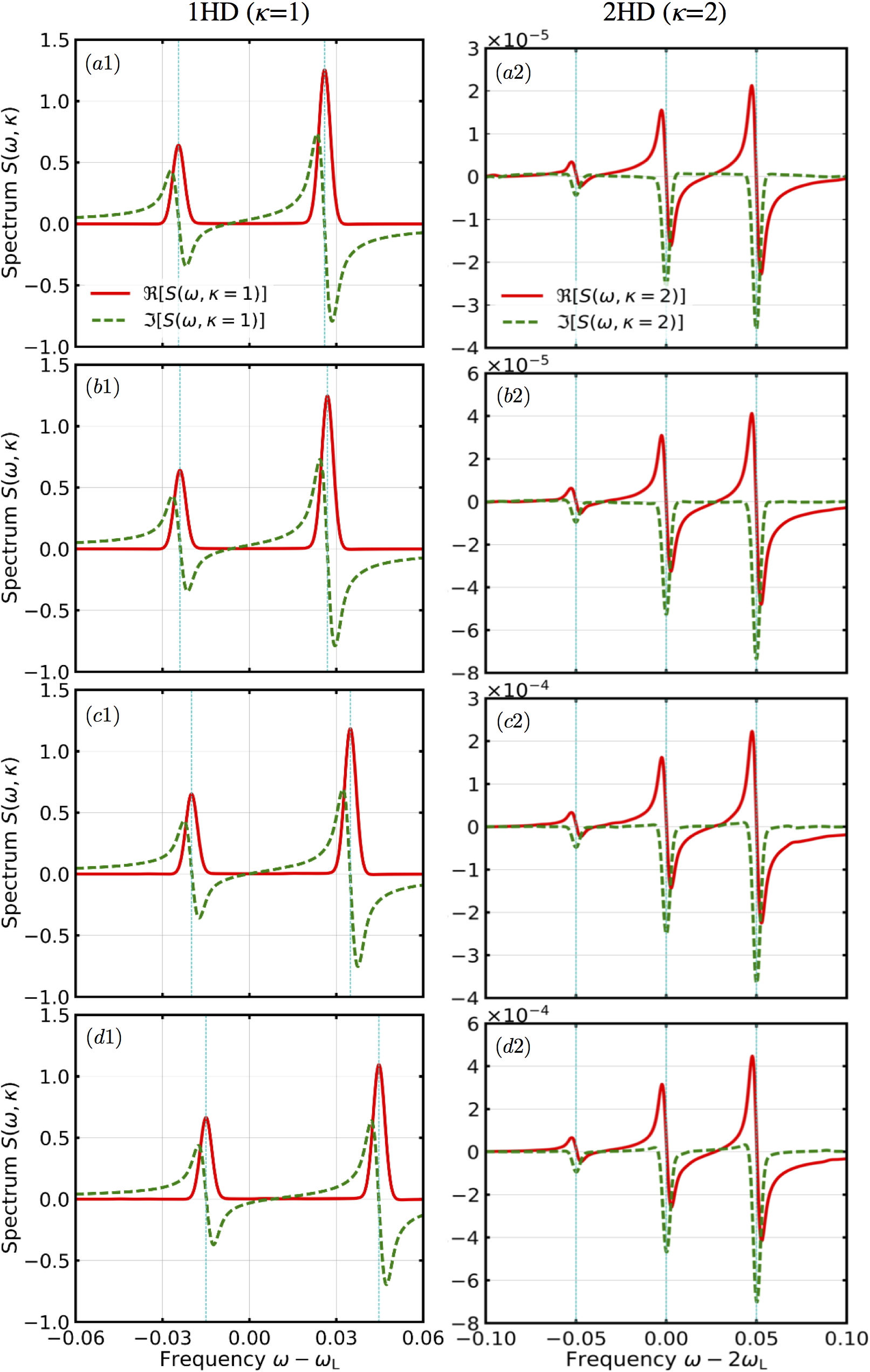}
\caption{(color online) The real (solid red line) and imaginary (dashed green line) parts of the 1HD (left column with $\kappa=1$) and 2HD (right column with $\kappa=2$) spectrum $S(\omega,\kappa)$ in Eq.~(\ref{eq:Swk}) for various dipole-dipole interactions. From top to bottom: $V_{{\rm ee}}=0.0005$, $0.001$, $0.005$, and $0.01$, and $V_{{\rm ff}}=1.974V_{{\rm ee}}$. The parameters of the pulse are  $E_1=E_2=0.00384$,  $\sigma=10.4$ ($\int A(t){\rm d}t=0.1$), and $\delta_{\rm eg}\equiv\omega_{\rm eg}-\omega_{\rm L}=-0.025$.
Note the different intensity scales in each panel.  All energies are in units of $\omega_0$.}
\label{fig:dipole_dipole_spectrum_positive}
\end{figure}

\begin{figure}
\centering
 \includegraphics[width=.98\columnwidth]{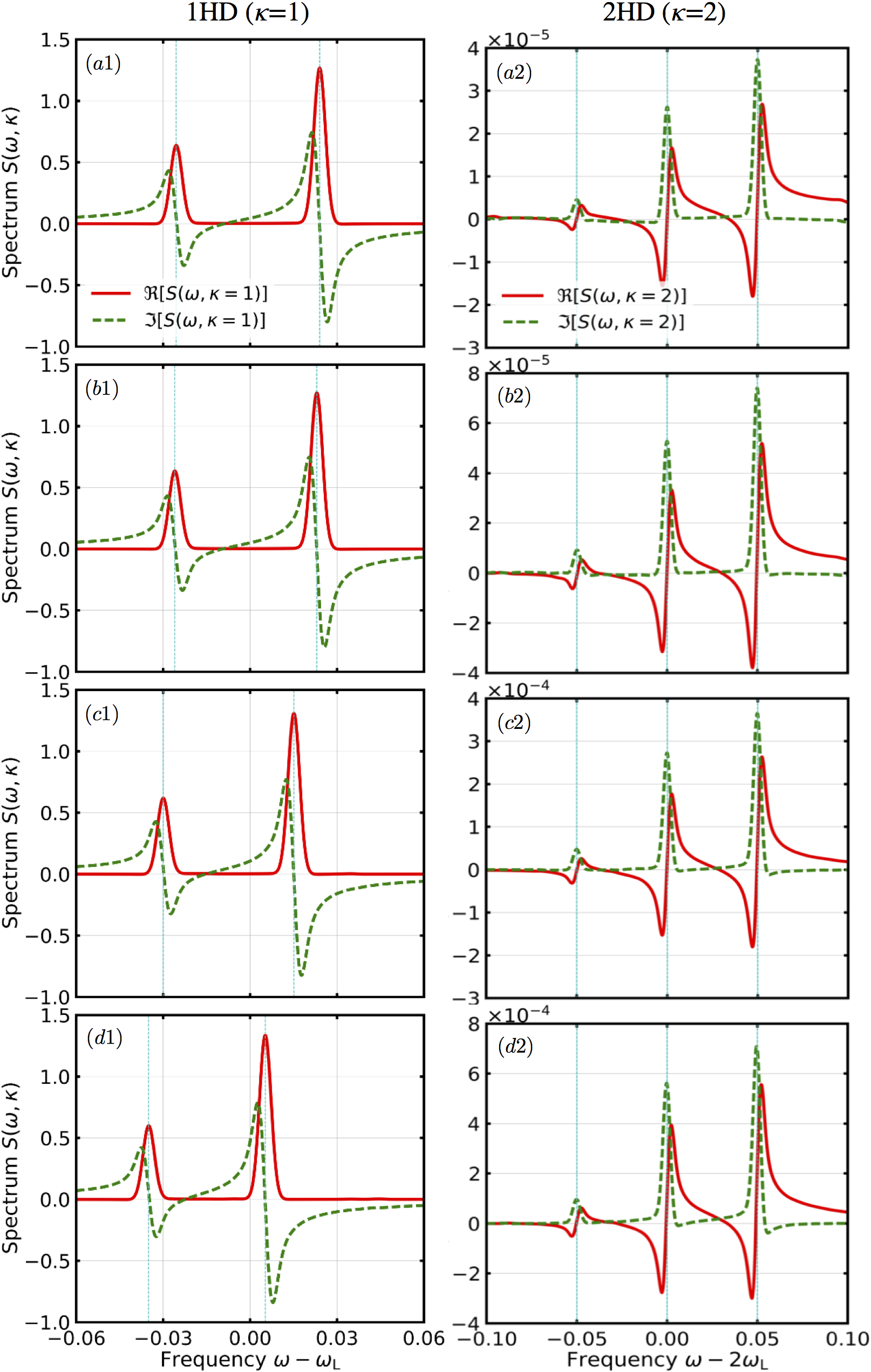}
\caption{(color online) Same as Fig.~\ref{fig:dipole_dipole_spectrum_positive}, but now for negative dipole-dipole interactions: $V_{\rm ee}=-0.0005$, $-0.001$, $-0.005$, and $-0.01$, and $V_{\rm ff}=1.974V_{\rm ee}$.}
\label{fig:dipole_dipole_spectrum_negative}
\end{figure}

In our fully numerical calculations we propagate the Hamiltonian specified in section~\ref{sec:model} (details on the numerical implementation can be found in the Supporting Information, section I.A).

In the following we consider two identical pulses, i.e. $\omega_1=\omega_2=\omega_{\rm L}$ and $E_1^{(0)}=E_2^{(0)}=E_0$ in a pulse pair.
The pulse envelope was taken  to be of Gaussian shape, i.e.
\begin{eqnarray}
A_j(t-t_j-\tau_m)=E_j^{(0)} e^{-{(t-t_j-\tau_m)^2}/{2\sigma^2}} . \label{eq:Gaussian_shape}
\end{eqnarray}
To understand the dependence on the parameters we varied the pulse width $\sigma$, the strength of the dipole-dipole interaction $V$, 
and the field strength $E_0$ of the pulses.
All parameters that were kept fixed for all calculations are given in  Table~\ref{table:pars}.
 The unit of energy we denote by $\omega_{0}$ (possible choices are for example bandwidth of the pulses or even $\omega_{0}=1/T_{\rm Rep}$).

For each parameter set we perform propagations for 4500 values of $t_{21}$ equally spaced between $t_{21}=0$ and $t_{21 }=2250$ and 1000 values of $\tau_{\rm m}$ with $T_{\rm Rep}=1$, resulting in the populations after the second pulse $P(t_{21},\tau_{\rm m}$).
Performing a numerical demodulation (described in the Supporting Information I.B), for each $t_{21}$ we have obtained a signal $\tilde{S}_{\rm F1uor}(t_{21})$.
This signal is then numerically Fourier transformed after multiplication with a Gaussian window function $e^{-{t^2_{21}}/{2 \bar{\sigma}^2}}$ with $\bar{\sigma}=500$ to avoid Fourier transform artefacts.
The final time and spacing of the $t_{21}$ values are chosen such that the peaks can be well resolved.
Larger times $t_{21}$ are preferable to reduce pulse overlap effects. 
The temporal FWHM of the window function is chosen to be $2\sqrt{2\ln 2}\bar{\sigma}\sim1177.41$ which results in a FWHM of $2\sqrt{2\ln 2}\frac{1}{\bar{\sigma}}\sim0.0047$ in frequency domain.
The resulting spectra exhibit sharp resonances which sit on a broad background.
We identified this background to be caused by pulse overlap effects.
In all results presented in the following, we have subtracted this background (details about the numerical procedure can be found in the Supporting Information, I.C).

Figs.~\ref{fig:dipole_dipole_spectrum_positive} and~\ref{fig:dipole_dipole_spectrum_negative} show example-spectra.
Each row shows for a set of parameters the real and imaginary parts of the $1$st harmonic demodulation (1HD) (left) and $2$nd harmonic demodulation (2HD) (right) spectrum.
From top to bottom the absolute value of the dipole-dipole interaction is increased. 
In Fig.~\ref{fig:dipole_dipole_spectrum_positive} all dipole-dipole interactions are positive and in Fig.~\ref{fig:dipole_dipole_spectrum_negative} they are negative.
The other parameters are specified in the figure caption. 
The vertical dashed lines indicate the peak positions obtained from the analytical treatment of section \ref{sec:analytics}.

\begin{figure}
\centering
  \includegraphics[width=.7\columnwidth]{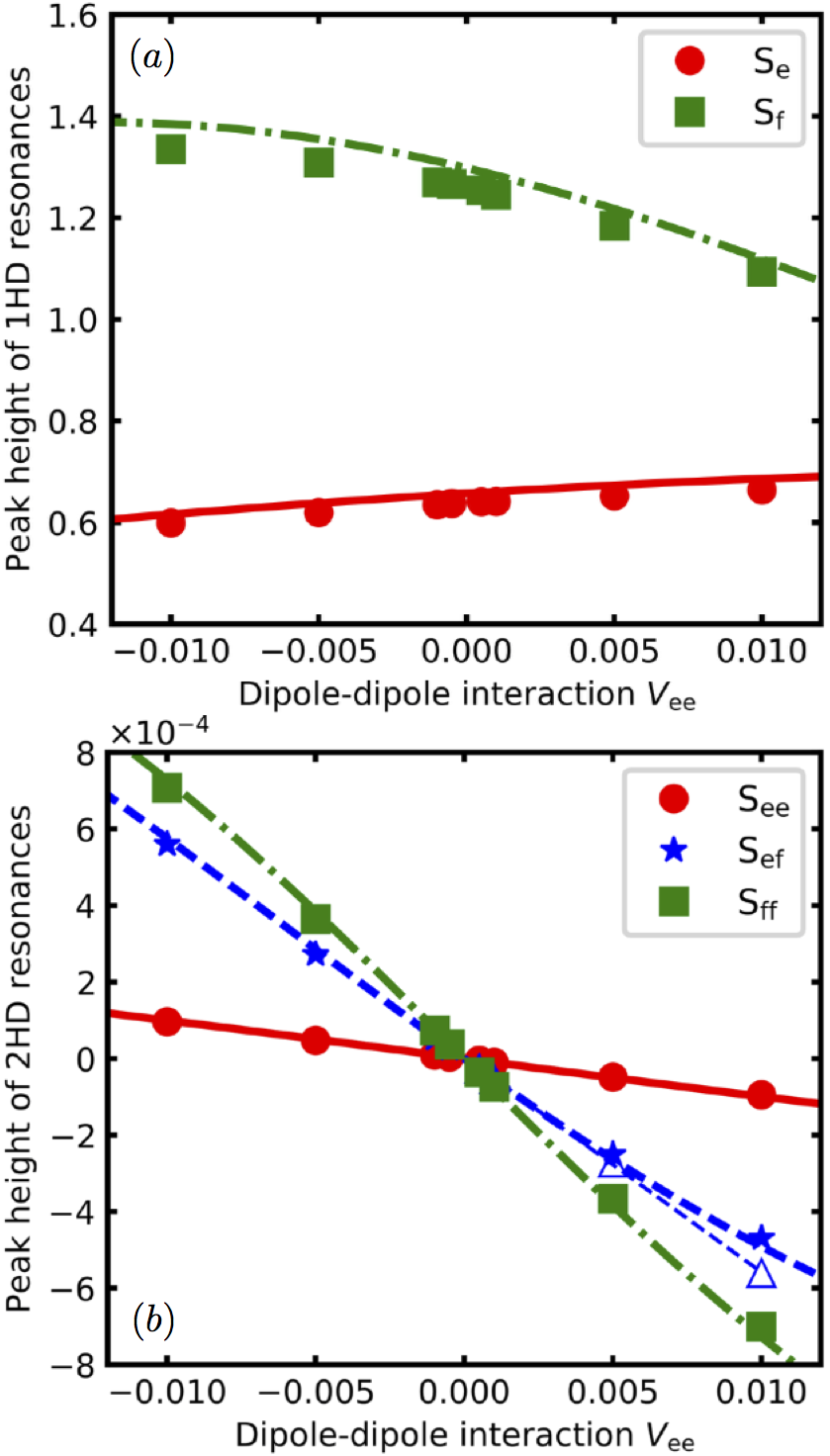}
\caption{(color online) Peak height (a) 1HD and (b) 2HD resonances corresponding respectively to (${\rm S_e}, {\rm S_f}$) and (${\rm S_{ee}}, {\rm S_{\rm ef}}, {\rm S_{\rm ff}}$) signals (see. e.g., Eqs.~(\ref{eq:D2_mod}), (\ref{eq:2D1_mod}), and (\ref{eq:D1D2_mod})). The filled circles, stars and squares are from numerical simulations while the solid, dashed and dot-dashed lines from the Fourier transformation of analytical results (i.e., $S^{(1)}(t_{21})$ and $S^{(2)}(t_{21})$ in Eqs.~(\ref{eq:1HD}) and (\ref{eq:2HD})) of section \ref{sec:analytics}. Other parameters are same as those in Fig.~\ref{fig:dipole_dipole_spectrum_positive}. The deviation between analytical and numerical calculations is caused by the approximation of non-overlapping pulses made in the analytical calculations. We have also performed calculations with shorter pulse length and found much better agreement. One example is shown in Appendix~\ref{sec:differenceAnalyticlNumeric}. 
The thin dashed blue line and the empty triangles emerge from the solid blue curve via the transformation $- V_{\rm ee}\rightarrow + V_{\rm ee}$ and a sign change of the peak amplitude. The curves for $S_{\rm ee}$ and $S_{\rm ff}$ are symmetric with respect to the origin.}
\label{fig:dipole_dipole_peakheight.png}
\end{figure}

We will refer to the two peaks in the 1HD spectra as ${\rm S_{e}}$ and ${\rm S_{f}}$, respectively (in Figs.~\ref{fig:dipole_dipole_spectrum_positive} and~\ref{fig:dipole_dipole_spectrum_negative} the ${\rm S_{e}}$ peak is located at a smaller frequency than that of the ${\rm S_{f}}$ peak). The three peaks in the 2HD spectra are denoted as ${\rm S_{ee}}$, ${\rm S_{ef}}$, and ${\rm S_{ff}}$ (in the order of increasing frequency).

Before we present results for a wide range of parameters, let us first briefly discuss some features that can be observed in Fig.~\ref{fig:dipole_dipole_spectrum_positive}.
 First note that the peak positions in the 1HD are shifted with respect to the position of the non-interacting particles by the energy of the dipole-dipole interaction.
In contrast, in the 2HD they are always at the constant frequencies $2 \omega_{\rm eg}$, $2 \omega_{\rm fg}$ and $\omega_{\rm eg}+\omega_{\rm fg}$. 
While in the 1HD peaks appear in the real part, they appear in the imaginary part of the 2HD and have a negative sign.
To obtain a similar behaviour as for the 1HD one has to multiply the 2HD by $e^{i\pi/2}$ for positive dipole-dipole coupling and by $e^{-i\pi/2}$ for negative dipole-dipole interaction.
The amplitude of each peak in 1HD decreases slightly ($\sim 10$ \%) upon increasing the dipole-dipole interaction.
Remarkably, in the 2HD the peak amplitude increases by an order of magnitude.
The last point is highlighted in Fig.~\ref{fig:dipole_dipole_peakheight.png} in more detail.
In this figure peak amplitudes are plotted as a function of the strength of the dipole-dipole interaction, now also including negative values.
The top panel shows 1HD and the bottom panel shows 2HD.
The filled circles, stars and squares are from numerical calculations, and the solid, dashed and dot-dahsed lines from the analytical results of section \ref{sec:analytics}.

The peaks at $2\omega_{\rm eg}$ and $2\omega_{\rm fg}$ have the same absolute value of the peak amplitude for the same absolute value of the dipole-dipole interaction strength.
However, the peak at $\omega_{\rm eg}+\omega_{\rm fg}$ has a different absolute value of the intensity when changing the sign of the dipole-dipole interaction (as indicated by the thin dashed blue line and the unfilled triangles).

\begin{figure}
\centering
  \includegraphics[width=.7\columnwidth]{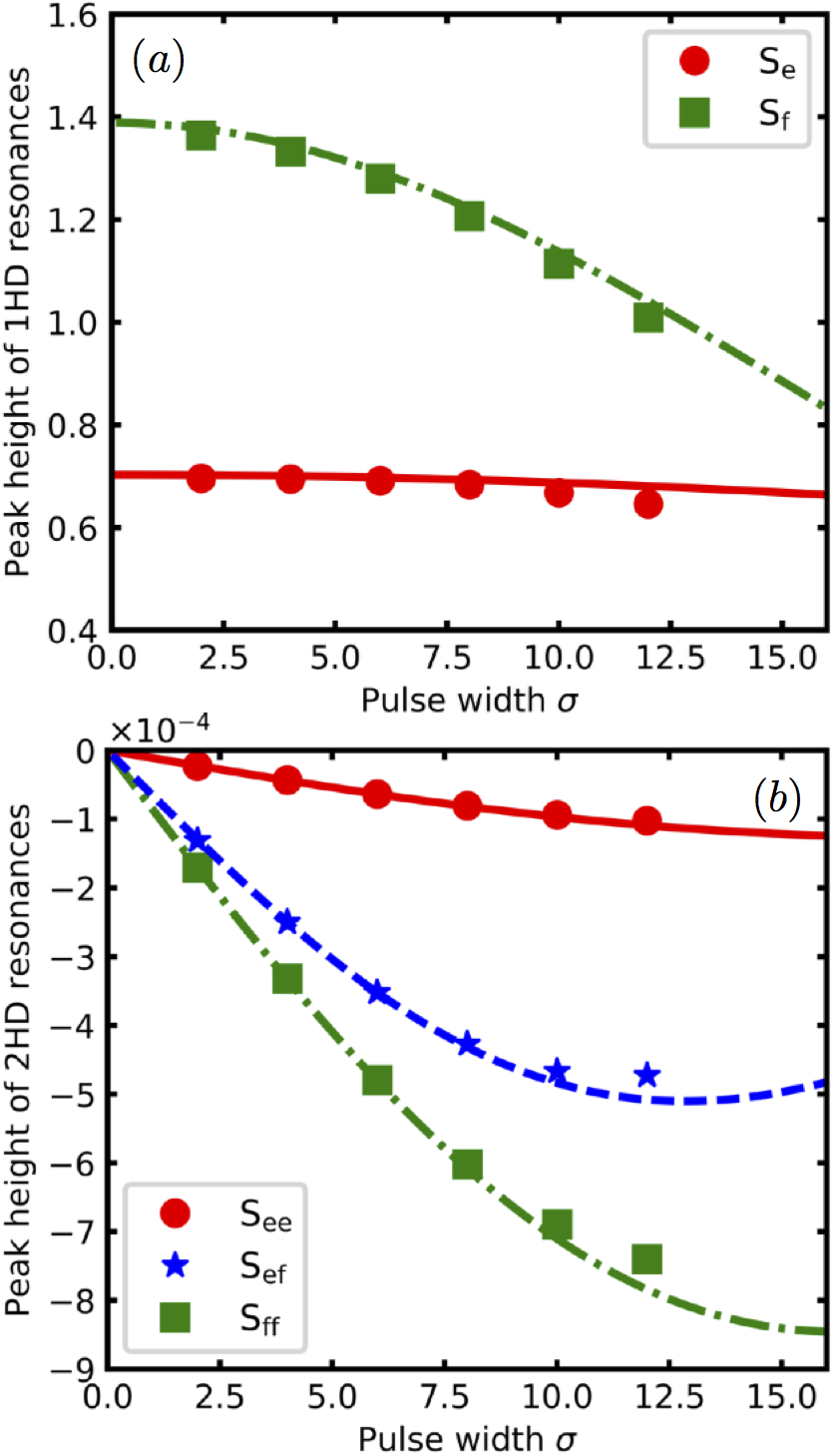}
\caption{(color online) Peak height of (a) 1HD and (b) 2HD resonances corresponding respectively to 1HD ($S_{\rm e}, S_{\rm f}$) and 2HD ($S_{\rm ee}, S_{\rm ef}, S_{\rm ff}$)   signals %(see. e.g., Eqs.~(\ref{eq:D2_mod}), (\ref{eq:2D1_mod}), and (\ref{eq:D1D2_mod})) 
versus laser pulse width. The filled circles, stars and squares given by harmonic spectrum $S(\omega,\kappa)$ in Eq.~(\ref{eq:Swk}) are from numerical simulations, while solid, dashed and dot-dashed lines are from the Fourier transformation of analytical results (i.e.,  $S^{(1)}(t_{21})$ and $S^{(2)}(t_{21})$ in Eqs.~(\ref{eq:1HD}) and (\ref{eq:2HD})) of section \ref{sec:analytics}. The dipole-dipole interaction is $V_{\rm ee}=0.01$ and all pulses are normalized to $\int A(t){\rm d}t=0.1$. The other parameters are the same as those in Fig.~\ref{fig:dipole_dipole_spectrum_positive}.}
\label{fig:pulse_width_peakheight}
\end{figure}

The effect of pulse width on the peak amplitude is shown in Fig.~\ref{fig:pulse_width_peakheight}.
Here we kept the `pulse-area' $I=\int A_j(t) {\rm d}t$ constant when changing the pulse width. 
For the used Gaussian pulses this implies a scaling of $E_0$ according to $E_0\sim 1/\sigma$. 
As expected the 1HD spectrum has a very weak dependence on the pulse width.
Essentially it reflects strength of the pulse-spectrum at the field strength of the respective transition.
The decrease at large pulse width comes from the fact that the spectral width of the pulse (which scales as $1/\sigma$) becomes comparable to the energetic separation between the transition frequency and the central frequency of the laser.
It is remarkable that the peak amplitude  in the 2HD has a very strong dependence on the pulse width.
In particular, for short pulse-width (i.e.\ large bandwidth of the pulse) the peak-amplitudes are decreasing for decreasing pulse-width.
This effect is related to the curvature (and magnitude) of the pulse-spectrum at the frequency of the relevant transitions. In addition it also depends in a non-trivial way on the difference between the pulse-spectrum at the `two-photon transition' (i.e.\ $\ket{{\rm gg}}\rightarrow \ket{{\rm ef}}$) and the  pulse-spectrum at the `one-photon transition'.
In the Appendix~\ref{sec:analyticRect} analytic calculations for rectangular pulses are performed in the perturbative regime with respect to the electric field (times transition dipole). From these results (in particular Eqs.~(\ref{eq:1st_comp_App})--(\ref{eq:2nd_12comp_App})) one sees that for the case when the detuning of the laser with respect to the single particle transitions is large compared to the dipole-dipole interaction, then the amplitude of the peaks consists of the product of two parts: one part leads to a `trivial' scaling (electric field $\times$ pulse width)$^n$, where $n=2$ for 1HD and $n=4$ for 2HD. The normalization of the electric field strength used in Fig.~\ref{fig:pulse_width_peakheight} keeps this factor constant. The second factor does for the 1HD signal in leading order not depend on the dipole-dipole interaction and approaches one for decreasing pulse width.
For the 2HD signal however, the leading term depends linearly on the dipole-dipole interaction (as we also found numerically in Fig.~\ref{fig:dipole_dipole_peakheight.png}).
The slope has non-trivial dependence on the pulse-width, and approaches zero for decreasing pulse-width. 
Let us note that often simulations of non-linear spectroscopy signals are performed  using delta pulses. 
From the foregoing considerations one sees that one has to be careful not to miss important features.

\begin{figure}
\centering
\includegraphics[width=0.7\columnwidth]{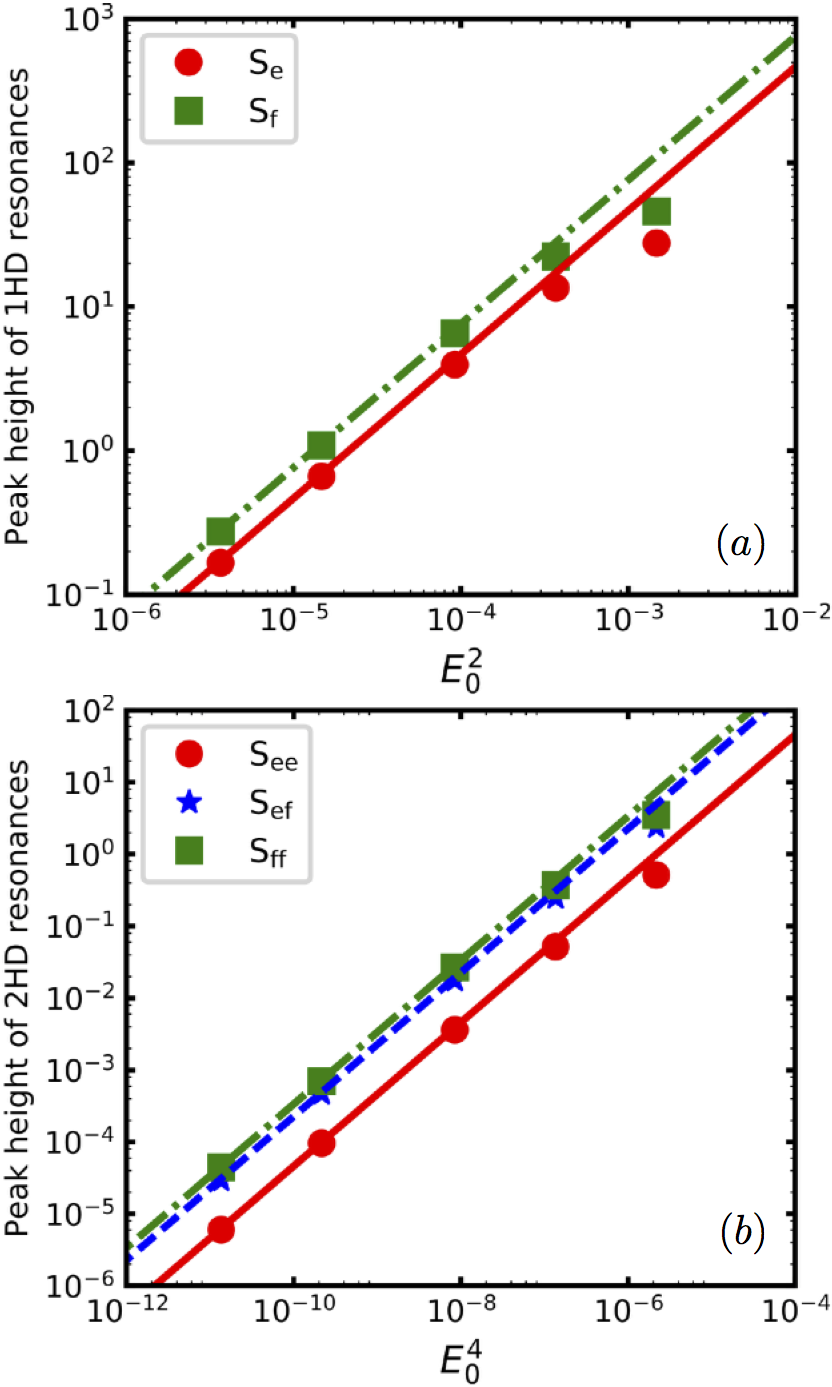} 
  \caption{(color online) Peak height of (a) 1HD and (b) 2HD resonances corresponding respectively to 1HD ($S_{\rm e}, S_{\rm f}$) and 2HD ($S_{\rm ee}, S_{\rm ef}, S_{\rm ff}$) signals %(see. e.g., Eqs.~(\ref{eq:D2_mod}), (\ref{eq:2D1_mod}), and (\ref{eq:D1D2_mod})) 
versus powers of the electric field strength. The filled circles, stars and squares given by harmonic spectrum $S(\omega,\kappa)$ in Eq.~(\ref{eq:Swk}) are from the fully numerical simulation (section \ref{sec:simulation}) while the solid, dashed and dot-dashed lines are given by the Fourier transformation of analytical results (i.e.,  $S^{(1)}(t_{21})$ and $S^{(2)}(t_{21})$ in Eqs.~(\ref{eq:1HD}) and (\ref{eq:2HD})) evaluated for Gaussian pulses. The dipole-dipole interaction is $V_{\rm ee}=0.01$ and all other pulse parameters such as the pulse width are kept constant. The other parameters are same as those in Fig.~\ref{fig:dipole_dipole_spectrum_positive}.}
\label{fig:electric_field_peakheight}
\end{figure}

Finally, in Fig.~\ref{fig:electric_field_peakheight} we show the dependence on the electric field strength.
The upper panel shows the 1HD peak intensities as a function of $E_0^2$ and the lower panel shows 2HD peak intensities as a function of $E_0^4$. 
The linear slope indicates that we are in the perturbative regime for field strength smaller than about $E_0\approx 0.01$. For larger values deviation from the linear behaviour becomes apparent.

\section{Perturbation theory with respect to the light-matter interaction \label{sec:analytics}}
In the following subsections we firstly present our general formulas for harmonic demodulated signals that are based on perturbation theory and applicable to arbitrary pulses. 
For Gaussian pulses  considered in our fully numerical simulation, our formulas together with a numerical integration of time-ordered double integrals show consistent results with the full numerical calculations and are displayed together with the numerical results.
In Appendix~\ref{sec:analyticRect} we apply our formulas to rectangular pulses for which fully analytical expressions can be obtained. These results are not directly comparable to the results of our numerical simulations but they help to develop physical understanding about the dependence of the signal on the various parameters. 

\subsection{General formulas}
In section II of the Supporting Information 
we provide information on our perturbative treatment and present intermediate results. Corresponding double-sided Feynman diagrams are provided in Appendix~\ref{sec:feynmandiag}. Here, in the main text, 
we will only provide final results after demodulation. 

Note that in the derivation of the formulas presented below, we have neglected  overlap between the two pulses.

\subsubsection{Some definitions}

It is convenient to define  for the pulses $j=1,2$ a single-interaction spectral amplitude
\begin{eqnarray}
\mathcal{A}_{j}(\omega) &=& \int_{-\infty}^{\infty} dt' A_j(t')  e^{i(\omega-\omega_j)t'}, \label{eq:single_pulse_spectrum}
\end{eqnarray}
and a double-interaction spectral amplitude 
\begin{eqnarray}
\mathcal{A}_{jj} (\omega, \tilde\omega) &=& \int_{-\infty}^{\infty} dt' \int_{-\infty}^{t'} dt''
A_j(t')  A_j(t'') \notag\\
&& \times e^{i(\omega-\tilde\omega-\omega_n)t'} \notag\\
&& \times e^{i(\tilde\omega-\omega_n)t''} .
\label{eq:double_pulse_spectrum}
\end{eqnarray}

Furthermore it is convenient to define the following `amplitudes':
 \begin{eqnarray}
A_{j}^{({\mathrm{g}\alpha},1)} = \mathcal{A}_j (\omega_{\alpha\mathrm{g}}+V_{\alpha\alpha}). \label{eq:ge_1}
\end{eqnarray} 
\begin{small}\begin{eqnarray}
A_{jj}^{(\alpha\alpha)} &=&  \mathcal{A}_{jj} (2\omega_{\alpha\mathrm{g}}, \omega_{\alpha\mathrm{g}}+V_{\alpha\alpha}),\\
A_{jj}^{(\mathrm{ef},\alpha)} &=&  \mathcal{A}_{jj} (\omega_{\mathrm{eg}}+\omega_{\mathrm{fg}}, \omega_{\mathrm{eg}}+V_{\alpha\alpha}) ,\\
A_{211}^{(\mathrm{ge},2)} \!\!&=&\!\! \mathcal{A}_2(\omega_{\mathrm{fg}}-V_{\mathrm{ee}}) \mathcal{A}_{11} (\omega_{\mathrm{eg}}+\omega_{\mathrm{fg}}, \omega_{\mathrm{eg}}+V_{\mathrm{ee}}), \label{eq:eg_ampl}\\
A_{211}^{(\mathrm{gf},\bar{2})} \!\!&=&\!\! \mathcal{A}_2(\omega_{\mathrm{eg}}-V_{\mathrm{ff}}) \mathcal{A}_{11} (\omega_{\mathrm{eg}}+\omega_{\mathrm{fg}}, \omega_{\mathrm{eg}}+V_{\mathrm{ee}}), \\
A_{211}^{(\mathrm{ge},\bar{2})} \!\!&=&\!\! \mathcal{A}_2(\omega_{\mathrm{fg}}-V_{\mathrm{ee}}) \mathcal{A}_{11} (\omega_{\mathrm{eg}}+\omega_{\mathrm{fg}}, \omega_{\mathrm{fg}}+V_{\mathrm{ff}}),\\
A_{211}^{(\mathrm{gf},2)} \!\!&=&\!\! \mathcal{A}_2(\omega_{\mathrm{eg}}-V_{\mathrm{ff}}) \mathcal{A}_{11} (\omega_{\mathrm{eg}}+\omega_{\mathrm{fg}}, \omega_{\mathrm{fg}}+V_{\mathrm{ff}}),\\
A_{211}^{(\mathrm{ge},3)} &=& \mathcal{A}_2(\omega_{\mathrm{eg}}-V_{\mathrm{ee}}) \mathcal{A}_{11} (2\omega_{\mathrm{eg}}, \omega_{\mathrm{eg}}+V_{\mathrm{ee}}),\\
A_{211}^{(\mathrm{gf},3)} &=& \mathcal{A}_2(\omega_{\mathrm{fg}}-V_{\mathrm{ff}}) \mathcal{A}_{11} (2\omega_{\mathrm{fg}}, \omega_{\mathrm{fg}}+V_{\mathrm{ff}}).
\label{eq:amplitudes_end}
\end{eqnarray}\end{small}

\subsubsection{The first harmonic demodulated signal}

The first harmonic demodulated signal is
\begin{equation}
S^{(1)}(t_{21})=S_{\rm e}(t_{21})+S_{\rm f}(t_{21}) \label{eq:1HD}
\end{equation}
with 
 \begin{eqnarray}
S_{\alpha}(t_{21})&=&\mu_{\alpha}^2 \Lambda_{\alpha} \cos(\Phi_{\alpha}(t_{21})), \label{eq:D2_mod}
\end{eqnarray} 
where  
\begin{eqnarray}
\Lambda_{\alpha} &=& A_{1}^{({\mathrm{g}\alpha},1)} A_{2}^{({\mathrm{g}\alpha},1)}, \label{eq:1st_comp} \\
\Phi_{\alpha} &=& (\omega_{\alpha\mathrm{g}}+V_{\alpha\alpha}-\omega_{\rm M})t_{21}. \label{eq:1st_phase}
\end{eqnarray} 

From the expression Eq.~(\ref{eq:1HD}) one sees that the 1HD spectrum consists of two contributions.
From Eq.~(\ref{eq:D2_mod}) together with Eq.~(\ref{eq:1st_phase}) one sees that each contribution results in a single peak located at the transition frequencies $\omega_{\alpha\mathrm{g}}+V_{\alpha\alpha}$ of the interacting particles ($\alpha={\rm e},\, {\rm f}$).
The intensity of these peaks is determined by Eq.~(\ref{eq:1st_comp}), which is a real function.

\subsubsection{The second harmonic demodulated signal}
The second harmonic demodulated signal is
\begin{equation}
S^{(2)}(t_{21})=S_{\rm ee}(t_{21})+S_{\rm ef}(t_{21})+S_{\rm ff}(t_{21}) \label{eq:2HD}
\end{equation}
with
 \begin{eqnarray}
S_{\alpha\alpha} &=& \frac{\mu_\alpha^4}{2}  \{ \Re[\Lambda_{\alpha\alpha}] \cos(\Phi_{\alpha\alpha})-\Im[\Lambda_{\alpha\alpha}] \sin(\Phi_{\alpha\alpha}) \},
\label{eq:2D1_mod}
\\
S_{\mathrm{ef}} \!\!&=&\!\! \frac{\mu_{\mathrm{e}}^2 \mu_{\mathrm{f}}^2}{4} \{ \Re[\Lambda_{\mathrm{ef}}] \cos(\Phi_{\mathrm{ef}}) -\Im[\Lambda_{\mathrm{ef}}] \sin(\Phi_{\mathrm{ef}}) \}.\label{eq:D1D2_mod}
\end{eqnarray} 
The coefficients and phases are given by
\begin{eqnarray}
\Lambda_{\alpha\alpha} &=& 2(A_{11}^{(\alpha\alpha)})^* A_{22}^{(\alpha\alpha)} -A_{2}^{(\mathrm{g}\alpha,1)} (A_{211}^{(\mathrm{g}\alpha,3)} )^*, \label{eq:2nd_comp}
\\
\Phi_{\alpha\alpha} &=& (2\omega_{\alpha\mathrm{g}}-2\omega_{\rm M} )t_{21},  \label{eq:2nd_phase}
\end{eqnarray}
and
\begin{eqnarray}
\Lambda_{\mathrm{ef}} &=& 2(A_{11}^{(\mathrm{ef},1)})^* A_{22}^{(\mathrm{ef},1)}
+2(A_{11}^{(\mathrm{ef},2)})^* A_{22}^{(\mathrm{ef},2)} \notag\\
&&+2A_{22}^{(\mathrm{ef},1)} (A_{11}^{(\mathrm{ef},2)})^*  +2 (A_{11}^{(\mathrm{ef},1)})^* A_{22}^{(\mathrm{ef},2)} \notag\\
&&-A_{2}^{(\mathrm{ge},1)} [(A_{211}^{(\mathrm{ge},2)})^* + (A_{211}^{(\mathrm{ge},\bar{2})})^* ]\notag\\
&&-A_{2}^{(\mathrm{gf},1)} [(A_{211}^{(\mathrm{gf},2)})^* +(A_{211}^{(\mathrm{gf},\bar{2})} )^*], \label{eq:2nd_12comp}\\
\Phi_{\mathrm{ef}} &=& (\omega_{\mathrm{eg}}+\omega_{\mathrm{fg}}-2\omega_{\rm M} )t_{21} . \label{eq:2ndCross_phase} 
\end{eqnarray}

From the expression Eq.~(\ref{eq:2HD}) one sees that the 2HD spectrum consists of three contributions.
These contributions correspond to peaks at two times the transition frequencies of the uncoupled particles (see Eq.~(\ref{eq:2D1_mod}) and Eq.~(\ref{eq:2nd_phase})) and to one peak that has a frequency of the sum of these transition frequencies (see Eq.~(\ref{eq:2ndCross_phase})).
The `intensity' of these peaks is determined by Eq.~(\ref{eq:2nd_comp}) and Eq.~(\ref{eq:2nd_12comp}), which are in general complex functions.

\subsubsection{Spectra in frequency domain}
The `spectrum'  can be obtained by applying the Fourier transformation, namely, 
$\mathcal{F}\{u(t)\cos(\omega_0 t)\}=\frac{1}{2}\sqrt{\frac{\pi}{2}}[\delta(\omega+\omega_0)+\delta(\omega-\omega_0)]-\frac{1}{2}\frac{i}{\sqrt{2\pi}}(\frac{1}{\omega+\omega_0}+\frac{1}{\omega-\omega_0})$ and $\mathcal{F}\{u(t)\sin(\omega_0 t)\}=\frac{i}{2}\sqrt{\frac{\pi}{2}}[\delta(\omega+\omega_0)-\delta(\omega-\omega_0)]+\frac{1}{2}\frac{1}{\sqrt{2\pi}}(\frac{1}{\omega+\omega_0}-\frac{1}{\omega-\omega_0})$ where $u(t)$ is the unit step function ($u(t)=1$ for $t\ge0$ and $0$ for $t<0$) and $\mathcal{F}\{x(t)\}=\frac{1}{\sqrt{2\pi}}\int_{-\infty}^{\infty} x(t) e^{-i\omega t} dt$.
The factor  $\omega_0$ has to be chosen according to the phases $\Phi_\alpha$, $\Phi_{\alpha\alpha}$ and $\Phi_{\rm ef}$ defined in Eqs.~(\ref{eq:1st_phase}), (\ref{eq:2nd_phase}) and (\ref{eq:2ndCross_phase}).

In order to compare with results given by the numerical simulation, the signals are further multiplied by the same Gaussian window function $e^{-{t_{21}^2}/{2\bar{\sigma}^2}}$ \footnote{By using the convolution theorem of Fourier transformation, we have $\Re[\mathcal{F}\{u(t)\cos(\omega_0 t) w(t)\}]=\frac{\bar{\sigma}}{4} [e^{-\frac{(\omega+\omega_0)^2\bar{\sigma}^2}{2}}+e^{-\frac{(\omega-\omega_0)^2\bar{\sigma}^2}{2}}]$ and $\Im[\mathcal{F}\{u(t)\sin(\omega_0 t) w(t)\}]=\frac{\bar{\sigma}}{4} [e^{-\frac{(\omega+\omega_0)^2\bar{\sigma}^2}{2}}-e^{-\frac{(\omega-\omega_0)^2\bar{\sigma}^2}{2}}]$. Here we used $w(t_{21})= e^{-\frac{t_{21}^2}{2\bar{\sigma}^2}}$ and $\mathcal{F}\{w(t_{21})\}=\bar{\sigma} e^{-\frac{\omega_0^2\bar{\sigma}^2}{2}}$.}.

\subsection{Gaussian pulses -- comparison to numerics\label{}}

The result for Gaussian pulses can be given by our formulas above together with a  numerical integration of time-ordered double integrals in Eq.~(\ref{eq:double_pulse_spectrum}).
The results are shown in Figs. \ref{fig:dipole_dipole_peakheight.png}, \ref{fig:pulse_width_peakheight}, and  \ref{fig:electric_field_peakheight} as solid, dashed and dot-dashed lines. One sees a very good agreement with our fully numerical simulation.
The reason for this good agreement is that the electric field strength of the pulses in the numerical simulation has been chosen to be in the perturbative regime.
This can be seen in Fig.~\ref{fig:electric_field_peakheight} where we present numerical calculations as a function of the field strength.
The linear dependence of the 1HD  with $E_0^2$ and the 2HD with $E_0^4$ indicates the validity of the perturbative treatment.
The solid, dashed and dot-dashed lines are again results from the analytical perturbative treatment.
One sees that only for the last two numerical points deviations from the perturbative calculations become visible.

 Deviations between analytical and numerical calculation are caused by the long pulse length used in the numerical simulations, where overlap of  the first and second pulse is fully included (while it is ignored in the analytical treatment). 
We have also performed numerical calculations with shorter pulse length and found much better agreement. One example is shown in Appendix~\ref{sec:differenceAnalyticlNumeric}.

\section{Applicability to molecular and atomic systems}
Although in the present study our focus is not on a particular system, we will in the following comment on two different systems.
The first one is the case of organic molecules in solution or on a substrate, the second one are atoms in a gas, similar to the situation of the experiments of Bruder et al.~\cite{BruderStienkemeier15PRA}. 

\subsection{Organic dye molecules}
One application of the present method is the investigation of dipole-dipole interacting organic dye molecules.
For such molecules, the electronic transition dipoles are typically in the order of several Debye. 
Dimers with specific interaction strength could be formed for example by linking the organic molecules by spacers \cite{ZhLiZh15_20108_,C5SC04956C}, by self-assembly in helium-nanodroplets \cite{WeSt03_125201_,RoEiDv11_054907_} or by depositing the molecules on a surface with a small coverage \cite{MPaMa13_064703_,MIzVl15_121408_} (using single molecule spectroscopy techniques one then can select close-by molecular pairs).
The distances between the molecules can range from a few {\AA}ngstr\"om to several nanometer. 
For close molecules the strength of the dipole-dipole interaction can be of the order of $\pm$100\, cm$^{-1}$.  
At a distance of 10\, nm one has interactions in the order of $\pm$0.1\, cm$^{-1}$.
While interactions of the order of $\pm$100\, cm$^{-1}$ lead to line shifts that can usually be well resolved by linear absorption spectroscopy, for interactions around $|V|\approx 1\; {\rm cm}^{-1}$ this becomes difficult because of line-broadening.  
Let us briefly relate these numbers to the present calculations.
Taking as unit of energy $\omega_0=100\,{\rm cm}^{-1}$, then in Figs.~\ref{fig:dipole_dipole_spectrum_positive}, \ref{fig:dipole_dipole_spectrum_negative} and \ref{fig:dipole_dipole_peakheight.png} the used pulse width $\sigma=10.4$ corresponds to $\approx 3\,{\rm ps}$ and the detuning $\omega_{\rm eg}-\omega_{\rm L}=-2.5\,{\rm cm}^{-1}$. 
For most molecules the electronically excited states are energetically well separated. 
Then the states $\ket{\rm e}$ and $\ket{\rm f}$ would correspond to vibronic states. 
If their energetic separation is much larger than the spectrum of the pulse, then essentially only one state contributes. 
However, for energetically very close energies (which is in particular the case for highly excited vibrational states) then two (or even more) states can contribute. The above example shows that one might be able to obtain information on the dipole-dipole coupling strength in the case when this interaction is weak compared to the energetic disorder of the individual molecules.

\subsection{Dilute potassium gas}
In the experiments of Ref.~\onlinecite{BruderStienkemeier15PRA}  a dilute gas ($10^{10}\,$cm$^{-3}$) of potassium atoms has been investigated.
In the resulting spectra at higher harmonic demodulation, peaks are visible at the energies that correspond to multiple excited particles. 
However, it is not clear how to model the resulting spectra (e.g. understand the intensities of the various peaks).
It has been suggested that the applied PM scheme reduces the experimental observable to an effective many-particle operator \cite{Mukamel2016JCP}. 
As a consequence, collective resonances should be observed independent of the presence of interactions among the atoms in the system. 
However, it is not obvious why the experimental procedure should result in a many-body measurement operator since the fluorescence is detected via a photodiode in the proportional range.

Although our model may neglect some aspects relevant in an atomic gas of many particles, we will nevertheless briefly relate the present results to the above mentioned experiment. 
 For a potassium atom, we make the following assignments:
 $|{\rm g}\rangle=|4 {}^2\!S_{1/2}\rangle$, $|{\rm e}\rangle=|4 {}^2\!P_{1/2}\rangle$, and $|{\rm f}\rangle=|4 {}^2\!P_{3/2}\rangle$.
In the 1HD one sees two peaks corresponding to the transition ${\rm g}\rightarrow{\rm e}$ and ${\rm g}\rightarrow{\rm f}$ at frequencies $\omega_{\rm eg}$ and $\omega_{\rm fg}$, respectively (for potassium one has $\frac{\mu_{\rm f}^2}{\mu_{\rm e}^2}\sim1.96$). 
The 2HD spectrum has peaks at $2\omega_{\rm eg}$, $2\omega_{\rm fg}$ and  $\omega_{\rm eg}+\omega_{\rm fg}$.
In the experiment, the magnitude of these peaks is approximately one order of magnitude smaller as the one of the 1HD peaks.

To obtain a rough prediction from our theoretical model we use experimental parameters for pulse intensities and width.
The dipole-dipole interaction we estimate from the mean particle separation in the gas.
Using these numbers we find that the 2HD is six orders of magnitude smaller than the 1HD. 
This discrepancy between our model and the experiment could have several reasons:
In the PM experiments Ref.~\onlinecite{BruderStienkemeier15PRA} high repetition rate lasers have been used.
The repetition rate is faster than the radiative lifetime of a single potassium atom. Thus the system may not be relaxed to the ground state before a consecutive pulse train perturbs the sample again. 
Such an effect is discussed in Ref.~\onlinecite{OsShHa16_053845_} for the case of a two-photon transition.
In the experiment not two atoms at a well-defined distance are addressed but the laser interacts with a large number of atoms in a gas. In our estimations we have assumed that the dominant contributions come from nearest neighbours and simply considered the average nearest neighbor distance. 
In a more detailed calculation one has to average over distances and relative orientations (also the assumption that the two atoms see the same field at a certain time is no longer correct, since the distance between the atoms is much larger than the wavelength). 
Note that depending on the relative orientation one can have positive and negative values of the dipole-dipole interaction.
Regarding the orientation of the transition dipoles some additional care is necessary. 
For atoms the transition dipoles are not simple vectors. 
And in particular for the dipole-dipole interaction one has to take the detailed level substructure into account (see e.g.\ the discussions in Refs.~\onlinecite{RoHeTo04_042703_} and~\onlinecite{MoeWueAt11_184011_}).
In our theoretical modelling we have ignored correlation effects in the emission (like stimulated emission or superradiance).

\section{Discussion and conclusion \label{sec:conclusion}}

 In this work, we have investigated systematically the collective excitations of particles probed via phase-modulated non-linear spectroscopy within the framework of a one-body measurement operator. 
We have focused on two particles that can exchange interaction via transition dipole-dipole interaction. The laser-pulses had frequencies overlapping with the transition frequencies of the individual particles.
We developed analytical formulas for the (de)modulated signals in terms of a single- and double-interaction pulse spectral amplitudes. We obtained analytical results for rectangular pulses and presented fully numerical simulation for Gaussian pulses. 
Below we summarize our main findings:

 The peak positions in the 1HD spectra are shifted with respect to the position of the non-interacting particles by the energy of the dipole-dipole interaction.
In contrast, in the 2HD they are always at the constant frequencies $2 \omega_{\rm eg}$, $2 \omega_{\rm fg}$ and $\omega_{\rm eg}+\omega_{\rm fg}$. 

The complex 1HD spectrum has the typical features of absorption and dispersion:
Peaks appear in the real part (of the complex spectrum) and the corresponding dispersion is in the imaginary part.
However, for the 2HD spectrum there is an important difference: Peaks appear in the imaginary part 
and dispersion-like features are in the real part.
If one wants to have a similar functional form as the 1HD feature (to interpret the signal as absorption and dispersion)  one has to multiply the 2HD by a phase-factor. This phase factor depends on the sign of the interaction; it is $e^{i\pi/2}$ for positive dipole-dipole coupling and by $e^{-i\pi/2}$ for negative dipole-dipole interaction.

For the amplitudes of the 2HD spectrum we found a linear dependence with respect to the strength of the dipole-dipole interaction.

We found a strong effect of the pulse-shape on the peak amplitude in the 2HD. In particular we found that it has a non-monotonic behaviour as a function of pulse width.
For delta-pulses the 2HD signal vanishes.

Let us briefly comment on some of our approximations.
In the present work we have neglected all decoherence effects. We expect that this is a good approximation, as long as the linewidth of the transitions of the individual atoms is small compared to the bandwidth of the pulses.
 In the present treatment all correlations in the field have been neglected. For example stimulated emission or Dicke-superradiance could have an effect on the observed fluorescence \cite{DorfmanMukamel13PhysRevA}. 

In the present work we have tried to follow the experimental procedure of Ref.~\onlinecite{BruderStienkemeier15PRA} as closely as possible.
However, instead of sampling with $\tau_{m}$ and performing a demodulation one can also employ a phase-cycling scheme, experimentally as well as numerically. 
In particular in the numerics this will give a considerable speed up and one can still treat non-perturbative effects and arbitrary pulse-shapes.
However, it will be difficult to treat effects specific to the demodulation procedure.
For example in the present numerical  approach one can model the lock-in amplifier in more detail or consider the effect of fast pulse repetition compared to the decay time of the particles.

%The peak shift in the 1HD spectrum with respect to the value of the uncoupled particles can indicate the coupling strength, as already shown in  Figs.~\ref{fig:dipole_dipole_spectrum_positive} and \ref{fig:dipole_dipole_spectrum_negative}. When the reference signal of the uncoupled system is unknown, it is also possible to extract the coupling strength from our proposed pulse excitation scheme. For example, the coupling strength would be obtained from the slope of the non-trivial peak height dependence of the 2HD spectrum on the small pulse width. This has been shown in Fig.~\ref{fig:pulse_width_peakheight} and also supported by analytical results e.g., Eqs.~(\ref{eq:D1122_taylor_pulsewidth}) and (\ref{eq:D1D2_taylor_pulsewidth}) for rectangle pulses. Note that the pulse width could be experimentally readily varied with a pulse shaper incorporated in the experimental setup.

Finally, let us note that our results indicate that phase modulation approach with higher order demodulation offers a possibility to detect weak interactions between particles when there is strong inhomogeneous broadening present. This is based on the fact that in the 2HD spectra dipole-dipole interaction manifests itself not in a line-shift (which is difficult to extract from a broadened spectrum), but from a peak amplitude. 

The coupling strength could be obtained from the non-trivial dependence of the peak amplitude of the 2HD spectrum on the pulse width, which is shown in Fig.~\ref{fig:pulse_width_peakheight} and also illustrated by analytical results e.g. Eqs.~(\ref{eq:D1122_taylor_pulsewidth}) and~(\ref{eq:D1D2_taylor_pulsewidth}) for rectangle pulses and small pulse width. The pulse width could be experimentally readily varied with a pulse shaper incorporated in the experimental setup.

\appendix

\section{Pulse length effect for the difference between analytics and numerics\label{sec:differenceAnalyticlNumeric}}

\begin{figure}[h]%[t]
\centering
  \includegraphics[width=.7\columnwidth]{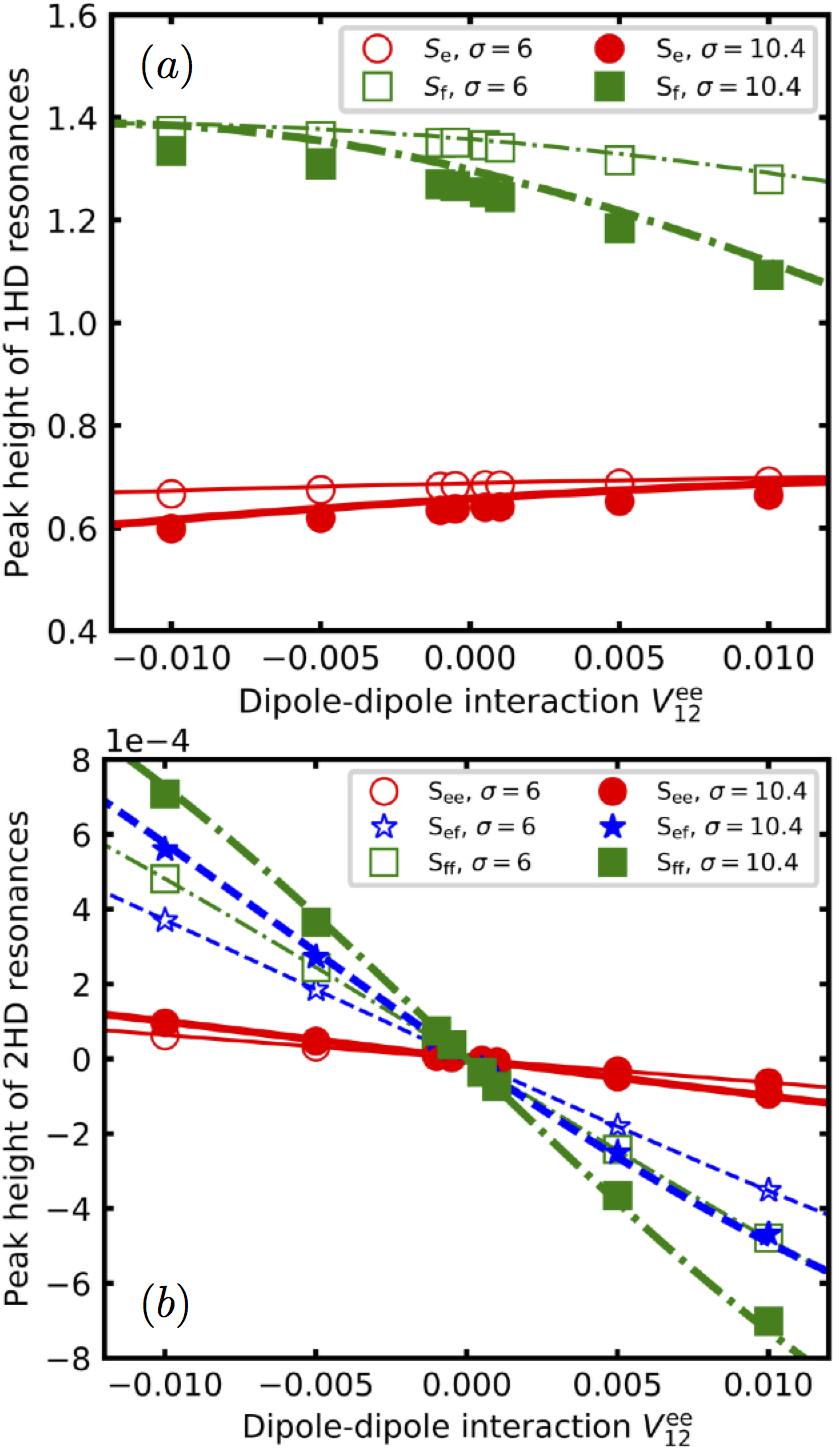}
\caption{(color online) Peak height (a) 1HD and (b) 2HD resonances corresponding respectively to (${\rm S_e}, {\rm S_f}$) and (${\rm S_{ee}}, {\rm S_{\rm ef}}, {\rm S_{\rm ff}}$) signals. The empty ($\sigma=6$) and filled ($\sigma=10.4$) circles, stars, and squares are from simulation while dashed ($\sigma=6$) and solid ($\sigma=10.4$) lines from the analytical results. Note that the results in Fig.~\ref{fig:dipole_dipole_peakheight.png} are reproduced here for comparison purposes.}
\label{fig:dipole_dipole_peakheight_som}
\end{figure}

In this appendix, we provide an example that shows how the agreement between our perturbation treatment and the full numerics depends on the length of the laser pulses. In Fig.~\ref{fig:dipole_dipole_peakheight.png}, it has been shown that there is deviation between analytical and numerical calculations. In this appendix, we perform calculations with shorter pulse length $\sigma=6$. A much better agreement shown in Fig.~\ref{fig:dipole_dipole_peakheight_som} supports our claim that the deviation in Fig.~\ref{fig:dipole_dipole_peakheight.png} is caused by the long pulse length $\sigma=10.4$ used in the numerical simulations.

\begin{figure}
\includegraphics[width=8.5cm]{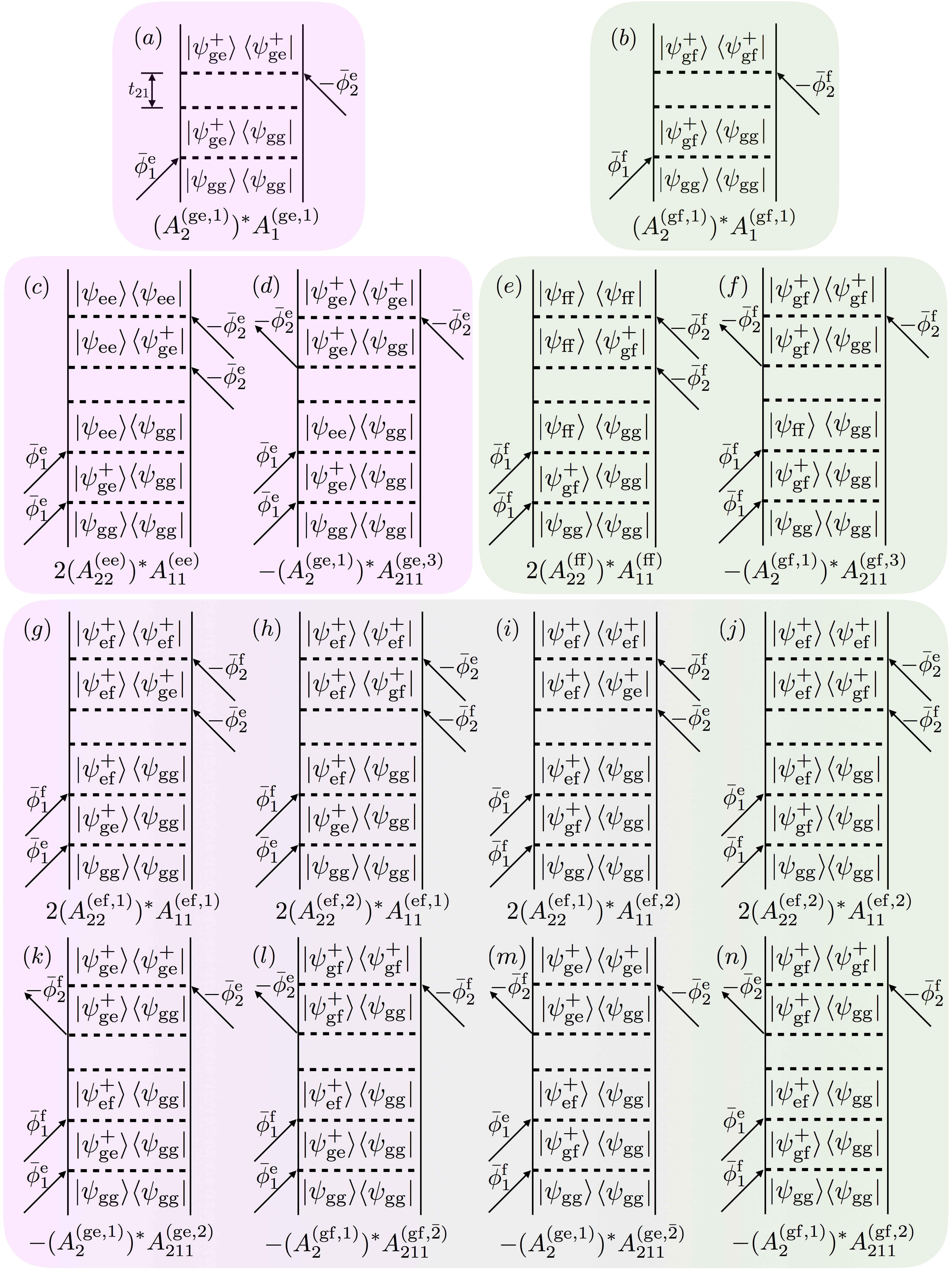}
\caption{\label{fig:Feynman}Double-sided Feynman diagrams that contribute to the first harmonic (a),(b) and second harmonic (c)-(n) demodulated signals. Below each diagram the resulting amplitude is given. A factor two indicates that in the final state there are two particles excited, i.e. two photons will be emitted. Here the phase is defined as $\phi_i^{jk\pm}=(\omega_{j\rm g}\pm V_{kk}-\omega_{\rm M})t_{i}$ for demodulated signals while it becomes $\phi_i^{jk\pm}=(\omega_{j\rm g}\pm V_{kk})t_i-(\Omega_i\tau_m+\phi_i^{(0)})$ ($i=1, 2$, $j={\rm e, f}$) before demodulation where only $\phi_i^{j\pm}(=\phi_i^{jj\pm})$ and $\phi_i^{jk-}(j\neq k)$ are involved in the Feynman diagrams.}
\label{fig:feynmandiag}
\end{figure}

\section{Fully analytical solutions for rectangular pulses\label{sec:analyticRect}}
In order to have a better understanding of physics behind and gain some insights from analytical results one would like to evaluate the various $\mathcal{A}$ of Eqs.~(\ref{eq:single_pulse_spectrum}) and (\ref{eq:double_pulse_spectrum}).
Due to time-ordered double integrals in Eq.~(\ref{eq:double_pulse_spectrum}), analytical solutions can only be obtained in some special cases.
Here we consider such a case where both pulses have a rectangle shape, namely,
\begin{equation}
A_j(t-t_j-\tau_m)=\left\{
\begin{array}{ccc}
E_0 &;& -\frac{\Delta}{2}< t < \frac{\Delta}{2}\\
& & \\
0 &;& {\rm {otherwise}}
\end{array}
\right.
\end{equation}
with $\Delta$ being pulse width.
 The single-interaction spectral amplitude defined in Eq.~(\ref{eq:single_pulse_spectrum}) becomes
\begin{eqnarray}
\mathcal{A}_{j}(\omega) &=& \mathcal{A} (\omega) = E_0\Delta \mathrm{sinc}\Big[(\omega-\omega_{\rm L}) \frac{\Delta}{2}\Big],
\label{eq:single_pulse_spectrum_rectangle}
\end{eqnarray}
where the sinc function is defined as $\mathrm{sinc}[x]=\frac{\sin x}{x}$.
The double-interaction spectral amplitude given by Eq.~(\ref{eq:double_pulse_spectrum}) becomes
\begin{eqnarray}
\mathcal{A}_{11} (\omega_{f}, \omega_i) \!\!&=&\!\! \frac{(E_0\Delta)^2}{i(\omega_i-\omega_{\rm L})}\Big\{\mathrm{sinc}\Big[\frac{(\omega_{ f}-2\omega_{\rm L})\Delta}{2}\Big] \notag\\
&&-\mathrm{sinc}\Big[\frac{(\omega_{ f}-\omega_i-\omega_L)\Delta}{2}\Big] e^{-i\frac{(\omega_i-\omega_{\rm L}) \Delta}{2}}\Big\}. \notag\\
\label{eq:double_pulse_spectrum_rectangle}
\end{eqnarray}
We want to emphasize that these two amplitudes, i.e. Eqs.~(\ref{eq:single_pulse_spectrum_rectangle}) and~(\ref{eq:double_pulse_spectrum_rectangle}) or their original definitions in Eqs.~(\ref{eq:single_pulse_spectrum}) and ~(\ref{eq:double_pulse_spectrum}) are key building blocks of harmonic signals of collective excitations.

It is important to note that the $\mathcal{A}_{jj}$ is in general complex-valued, while  $\mathcal{A}_{j}$ is always real. 

The expression Eq.~(\ref{eq:double_pulse_spectrum_rectangle}) is still quite complicated.
In the following, we perform series expansions to consider a regime of weak dipole-dipole interaction and also investigate the effect of finite duration of pulses.

\subsubsection{Dipole-dipole interactions small compared to detuning}
\label{sec:small-dipdip}

Here we consider the case $|\frac{V_{\rm ee}}{\delta_{\rm eg}}|\ll 1$ and $|\frac{V_{\rm ff}}{\delta_{\rm fg}}|\ll 1$ where the detunings with respect to the laser central frequency are defined as $\delta_{\rm eg}=\omega_{\rm eg}-\omega_{\rm L}$ or $\delta_{\rm fg}=\omega_{\rm fg}-\omega_{\rm L}$. 

Then Eq.~(\ref{eq:1st_comp}) for the first harmonic signals becomes
\begin{eqnarray}
\label{eq:1st_comp_App}
\Lambda_{\alpha} &\approx& (E_0\Delta)^2 \mathrm{sinc}^2 (\frac{\Delta}{2} \delta_{\alpha\rm g})\notag \\
&&-2 (E_0\Delta)^2  \frac{2-2\cos(\Delta\delta_{\alpha\rm g})-\Delta\delta_{\alpha\rm g}\sin(\Delta\delta_{\alpha\rm g})}{\Delta^2\delta_{\alpha\rm g}^2} \frac{ V_{\alpha\alpha}}{\delta_{\alpha\rm g}}, \notag \\ 
\end{eqnarray}
The second harmonic ones, Eqs.~(\ref{eq:2nd_comp}) and (\ref{eq:2nd_12comp}) become
\begin{eqnarray}
\label{eq:2nd_comp_App}
\Lambda_{\alpha\alpha} &\approx& -i (E_0\Delta)^4 \frac{\mathrm{sinc}^2(\frac{\Delta}{2}\delta_{\alpha\rm g}) [ 1-\mathrm{sinc}(\Delta\delta_{\alpha\rm g}) ]}{\Delta\delta_{\alpha\rm g}}
\frac{ V_{\alpha\alpha} }{\delta_{\alpha\rm g}},  \\
%\end{eqnarray}
%\begin{eqnarray}
\Lambda_{\rm ef} &\approx& -i (E_0\Delta)^4 \frac{\mathrm{sinc}(\frac{\Delta}{2}\delta_{\rm eg}) \mathrm{sinc}(\frac{\Delta}{2}\delta_{\rm fg})}{\Delta \sqrt{\delta_{\rm eg} \delta_{\rm fg}} }
 \mathcal{C} \frac{V_{\rm ee}+ V_{\rm ff}}{\sqrt{ \delta_{\rm eg} \delta_{\rm fg}} }, %\notag\\
\label{eq:2nd_12comp_App}
\end{eqnarray}
with
\begin{eqnarray}
 \mathcal{C} &=& 
 2 \cos[\frac{\Delta}{2} (\delta_{\rm fg}-\delta_{\rm eg})] 
 +\frac{2(\delta_{\rm fg}-\delta_{\rm eg})}{\Delta \delta_{\rm eg} \delta_{\rm fg} } 
 \sin[\frac{\Delta}{2} (\delta_{\rm fg}-\delta_{\rm eg})] \notag\\
 &&
 -\frac{\delta_{fg}^2+\delta_{\rm eg}^2}{\delta_{\rm eg} \delta_{\rm fg}}
 \mathrm{sinc}[\frac{\Delta}{2} (\delta_{\rm fg}+\delta_{\rm eg})] . 
\end{eqnarray}
The intensity of  collective resonances scales linearly with the dipole-dipole interaction.
 Collective signals become zero when the dipole-dipole interaction is absent.

Note that the lowest order term of $\Lambda_{\rm e}$ and $\Lambda_{\rm f}$ is real and positive, while for $\Lambda_{\rm ee}$, $\Lambda_{\rm ff}$, and $\Lambda_{\rm ef}$, they are  positive and imaginary. After inserting $\Lambda_{\rm ee}$, $\Lambda_{\rm ff}$, and $\Lambda_{\rm ef}$ into $S_{\rm ee}$, $S_{\rm ff}$, and $S_{\rm ef}$ in Eqs.~(\ref{eq:2D1_mod}) and~(\ref{eq:D1D2_mod}) respectively, the second harmonic demodulated signals given by the demodulation of $S_{\rm ee}$, $S_{\rm ff}$, and $S_{\rm ef}$ have a phase shift $\frac{3\pi}{2}$ or $\frac{\pi}{2}$ for positive and negative dipole-dipole interaction respectively, from the first harmonic demodulated signal (i.e. phase demodulation of Eq.~(\ref{eq:D2_mod})). 
This phase shift that does not appear when considering delta pulses, has also been observed in our simulation in Figs.~\ref{fig:dipole_dipole_spectrum_positive}, \ref{fig:dipole_dipole_spectrum_negative}, and \ref{fig:dipole_dipole_peakheight.png}.

\subsubsection{Dipole-dipole interaction small compared to bandwidth of pulse}

Here we consider $ |V_{\rm ee}\Delta|\ll1$ or $ |V_{\rm ff}\Delta|\ll1$ . 
Up to the third order, we have 
\begin{eqnarray}
\Lambda_\alpha &\approx& (E_0\Delta)^2 \Big(1 -\frac{1}{12} \Delta^2(\delta_{\alpha \rm g}+V_{\alpha\alpha})^2\Big), \label{eq:D1_taylor_pulsewidth} \\
\Lambda_{\alpha\alpha} &\approx& -i\frac{1}{6}   (E_0\Delta)^4 (V_{\alpha\alpha}\Delta) \Big(1+\frac{i}{3} (V_{\alpha\alpha} \Delta) \notag \\
&&\phantom{dddd}-\frac{6}{45}  [(V_{\alpha})^2 +(\delta_{\alpha\rm g})^2] \Delta^2\Big), \label{eq:D1122_taylor_pulsewidth}\\
\Lambda_{\rm ef} &\approx& -i\frac{1}{3}(E_0\Delta)^4(V_{\rm ee}+V_{\rm ff})\Delta\Big[1 +\frac{i}{6} (V_{\rm ee}+V_{\rm ff})  \Delta \notag\\
&&-\frac{1}{120} \mathcal{B} \Delta^2\Big], \label{eq:D1D2_taylor_pulsewidth}
\end{eqnarray} 
where the detunings with respect to the laser central frequency are again defined as $\delta_{\rm eg}=\omega_{\rm eg}-\omega_{\rm L}$ or $\delta_{\rm fg}=\omega_{\rm fg}-\omega_{\rm L}$ and are also taken to be small.
The expression for $\mathcal{B}$ is
\begin{eqnarray}
\mathcal{B} &=& 11[(V_{\rm ee})^2+(V_{\rm ff})^2+\omega_{\rm eg}^2+\omega_{\rm fg}^2] \notag\\
&&-6(V_{\rm ee} V_{\rm ff}+\omega_{\rm eg}\omega_{\rm fg})  \notag \\
&&+14(V_{\rm ff}-V_{\rm ee}) (\omega_{\rm fg}-\omega_{\rm eg})  \notag \\
&&+16\omega_{\rm L} (\omega_{\rm L}-\omega_{\rm eg}-\omega_{\rm fg}) .
\end{eqnarray}
As in the previous subsection, the intensity of collective resonances scales in lowest order linearly with the dipole-dipole interaction and collective signals become zero when the dipole-dipole interaction is absent.
Also the second harmonic demodulated signals given by the demodulation of $S_{\rm ee}$, $S_{\rm ff}$, and $S_{\rm ef}$ have a phase shift $\frac{3\pi}{2}$ or $\frac{\pi}{2}$ for positive and negative dipole-dipole interaction respectively, from the first harmonic demodulated signal. This disappears for delta-pulses.

It is interesting to note that  the imaginary part of $\Lambda_{\alpha\alpha}$, which gives rise to the absorptive features in the spectrum, is an odd function of $V_{\alpha\alpha}$. Therefor the  intensities for positive and negative dipole-dipole interaction  $|V_{\alpha\alpha}|$ have the   same magnitude but different signs. 
This is not the case for $\Lambda_{\rm ef}$.
This is because of the third order term containing $\mathcal{B} $. One finds that the absolute value of signal for $ V_{\alpha\alpha}<0 $ becomes larger than that for $ V_{\alpha\alpha}>0 $. This is exactly the phenomenon that we have observed in simulation for Gaussian pulses.

\section{Relation to double-sided Feynman diagrams\label{sec:feynmandiag}}

The origin of the theoretical formulas of section~\ref{sec:analytics} can be easily understood by considering the relevant double-sided Feynman diagrams using the general rules provided e.g.\ in Ref.~\onlinecite{Mukamel95}.
In Fig.~\ref{fig:feynmandiag} all diagrams are displayed that result in a phase $\Omega_{21} \tau_m$ (a and b) and $2\Omega_{21} \tau_m$, and therefore contribute to the 1HD and 2HD signal, respectively.
Below each diagram the resulting amplitude is provided.
Note that the appearance of a factor two in front of a diagram reflects the fact that in the final state two particles are excited and  consequently will emit two photons.


\begin{thebibliography}{10}
\providecommand{\url}[1]{\texttt{#1}}
\providecommand{\urlprefix}{URL }
\expandafter\ifx\csname urlstyle\endcsname\relax
  \providecommand{\doi}[1]{doi:\discretionary{}{}{}#1}\else
  \providecommand{\doi}{doi:\discretionary{}{}{}\begingroup
  \urlstyle{rm}\Url}\fi
\providecommand{\eprint}[2][]{\url{#2}}

\bibitem{StoneCundiffNelson09Science}
K.~W. Stone, K.~Gundogdu, D.~B. Turner, X.~Li, S.~T. Cundiff, and K.~A. Nelson,
  Science \textbf{324}, 1169 (2009).

\bibitem{KaraiskajCundiff10PRL}
D.~Karaiskaj, A.~D. Bristow, L.~Yang, X.~Dai, R.~P. Mirin, S.~Mukamel, and S.~T.
  Cundiff, Phys. Rev. Lett. \textbf{104}, 117401 (2010).

\bibitem{TurnerNelson10Nature}
D.~B. Turner and K.~A. Nelson, Nature \textbf{466}, 1089 (2010).

\bibitem{DaiCundiff12PRL}
X.~Dai, M.~Richter, H.~Li, A.~D. Bristow, C.~Falvo, S.~Mukamel, and S.~T.
  Cundiff, Phys. Rev. Lett. \textbf{108}, 193201 (2012).

\bibitem{GaoCundiffLi16}
F.~Gao, S.~T. Cundiff, and H.~Li, Opt. Lett. \textbf{41}, 2954 (2016).

\bibitem{TekavecMarcus07JCP}
P.~F. Tekavec, G.~A. Lott, and A.~H. Marcus, J. Chem. Phys. \textbf{127}, 214307
  (2007).

\bibitem{TekavecMarcus06JCP}
P.~F. Tekavec, T.~R. Dyke, and A.~H. Marcus, J. Chem. Phys. \textbf{125}, 194303
  (2006).

\bibitem{NardinCundiff13}
G.~Nardin, T.~M. Autry, K.~L. Silverman, and S.~T. Cundiff, Opt. Express
  \textbf{21}, 28617 (2013).

\bibitem{BruderStienkemeier2015PCCP}
L.~Bruder, M.~Mudrich, and F.~Stienkemeier, Phys. Chem. Chem. Phys. \textbf{17},
  23877 (2015).

\bibitem{BruderStienkemeier15PRA}
L.~Bruder, M.~Binz, and F.~Stienkemeier, Phys. Rev. A \textbf{92}, 053412 (2015).

\bibitem{WeSt03_125201_}
M.~Wewer and F.~Stienkemeier, Phys. Rev. B \textbf{67}, 125201 (2003).

\bibitem{FiSeEn08_12858_}
R.~F. Fink, J.~Seibt, V.~Engel, M.~Renz, M.~Kaupp, S.~Lochbrunner, H.-M. Zhao,
  J.~Pfister, F.~W\"{u}rthner, and B.~Engels, J. Am. Chem. Soc. \textbf{130}, 12858
  (2008).

\bibitem{Ei07_321_}
A.~Eisfeld, Chem. Phys. Lett. \textbf{445}, 321 (2007).

\bibitem{GuZuCh09_154302_}
J.~Guthmuller, F.~Zutterman, and B.~Champagne, J. Chem. Phys. \textbf{131},
  154302 (2009).

\bibitem{WoMo09_15747_}
J.~M. Womick and A.~M. Moran, J. Phys. Chem. B \textbf{113}, 15747 (2009).

\bibitem{DuNaPr15_072002_}
H.-G. Duan, P.~Nalbach, V.~I. Prokhorenko, S.~Mukamel, and M.~Thorwart, New J.
  Phys. \textbf{17}, 072002 (2015).

\bibitem{May-Kuehn}
V.~May and O.~K\"{u}hn, \emph{Charge and Energy Transfer Dynamics in Molecular
  Systems} (3rd, Wiley-VCH, 2011).

\bibitem{KaKrMa16_015504_}
K.~J. Karki, L.~Kringle, A.~H. Marcus, and T.~Pullerits, J. Opt. \textbf{18},
  015504 (2016).

\bibitem{TianWarren2002OptLett}
P.~Tian and W.~S. Warren, Opt. Lett. \textbf{27}, 1634 (2002).

\bibitem{Mukamel95}
S.~Mukamel, \emph{Principles of nonlinear optical spectroscopy} (Oxford University Press, 1995).

\bibitem{ZhLiZh15_20108_}
J.~Zhao, Y.~Li, J.~Zhang, L.~Zhang, J.~Y.~L. Lai, K.~Jiang, C.~Mu, Z.~Li,
  C.~L.~C. Chan, A.~Hunt, S.~Mukherjee, H.~Ade, X.~Huang, and H.~Yan, J. Mater.
  Chem. A \textbf{3}, 20108 (2015).

\bibitem{C5SC04956C}
P.~E. Hartnett, H.~S. S.~R. Matte, N.~D. Eastham, N.~E. Jackson, Y.~Wu, L.~X.
  Chen, M.~A. Ratner, R.~P.~H. Chang, M.~C. Hersam, M.~R. Wasielewski, and T.~J.
  Marks, Chem. Sci. \textbf{7}, 3543 (2016).

\bibitem{RoEiDv11_054907_}
J.~Roden, A.~Eisfeld, M.~Dvo\v{r}\'ak, O.~B\"unermann, and F.~Stienkemeier, J.
  Chem. Phys. \textbf{134}, 054907 (2011).

\bibitem{MPaMa13_064703_}
M.~M\"{u}ller, A.~Paulheim, C.~Marquardt, and M.~Sokolowski, J. Chem. Phys.
  \textbf{138}, 064703 (2013).

\bibitem{MIzVl15_121408_}
M.~M\"uller, S.~Izadnia, S.~M. Vlaming, A.~Eisfeld, A.~LaForge, and
  F.~Stienkemeier, Phys. Rev. B \textbf{92}, 121408 (2015).

\bibitem{Mukamel2016JCP}
S.~Mukamel, J. Chem. Phys. \textbf{145}, 041102 (2016).

\bibitem{OsShHa16_053845_}
V.~A. Osipov, X.~Shang, T.~Hansen, T.~Pullerits, and K.~J. Karki, Phys. Rev. A
  \textbf{94}, 053845 (2016).

\bibitem{RoHeTo04_042703_}
F.~Robicheaux, J.~V. Hern\'{a}ndez, T.~Top\c{c}u, and L.~D. Noordam, Phys. Rev.
  A \textbf{70}, 042703 (2004).

\bibitem{MoeWueAt11_184011_}
S.~M\"{o}bius, S.~W\"{u}ster, C.~Ates, A.~Eisfeld, and J.-M. Rost, J. Phys. B
  \textbf{44}, 184011 (2011).

\bibitem{DorfmanMukamel13PhysRevA}
K.~E. Dorfman and S.~Mukamel, Phys. Rev. A \textbf{87}, 063831 (2013).

\end{thebibliography}
\end{document}